\documentclass[twocolumn]{aastex62}

\received{December 31, 2018}
\accepted{April 15, 2019}

\usepackage{amsmath,amssymb}
\usepackage{comment}
\usepackage{mathrsfs}
\usepackage{courier}

\received{December 31, 2018}

\submitjournal{ApJ}

\begin{document}
\title{Constraining massive star activities in the final years through properties of supernovae and their progenitors}

\correspondingauthor{Ryoma Ouchi}
\email{ouchi@kusastro.kyoto-u.ac.jp}
\affiliation{Department of Astronomy, Kyoto University, Kitashirakawa-Oiwake-cho, Sakyo-ku, Kyoto 606-8502, Japan}
\author{Ryoma Ouchi}
\affiliation{Department of Astronomy, Kyoto University, Kitashirakawa-Oiwake-cho, Sakyo-ku, Kyoto 606-8502, Japan}

\author{Keiichi Maeda}
\affiliation{Department of Astronomy, Kyoto University, Kitashirakawa-Oiwake-cho, Sakyo-ku, Kyoto 606-8502, Japan}



\begin{abstract}
  Recent observations of supernovae (SNe) just after the explosion suggest that a good fraction of SNe have the confined circumstellar material (CSM) in the vicinity, and the pre-SN enhanced mass loss may be a common property. The physical mechanism of this phenomenon is still unclarified, and the energy deposition into the envelope has been proposed as a possible cause of the confined CSM. In this work, we have calculated the response of the envelope to various types of sustained energy deposition starting from a few years before the core collapse. We have further investigated how the resulting progenitor structure would affect appearance of the ensuing supernova.
  While it has been suspected that a super-Eddington energy deposition may lead to a strong and/or eruptive mass loss to account for the confined CSM, we have found that a highly super-Eddington energy injection into the envelope changes the structure of the progenitor star substantially, and the properties of the resulting SNe become inconsistent with usual SNe. This argument constrains the energy budget involved in the possible stellar activity in the final years to be at most one order of magnitude higher than the Eddington luminosity. Such an energy generation however would not dynamically develop a strong wind in the time scale of a few years. We therefore propose that a secondary effect (e.g., pulsation or binary interaction) triggered by the moderate envelope inflation, which is caused by sub-Eddington energy injection, likely induces the mass loss.
\end{abstract}



\section{Introduction} \label{sec:intro}

The evolution of massive stars ($M \gtrsim 8 M_{\odot}$) just a few years before the core collapse, which sets the initial condition for the ensuing supernova (SN), seems to be much more uncertain than previously believed. Recently, evidence has been accumulating that some massive stars experience the enhanced mass loss ($\dot{M} \gtrsim 10^{-4} M_{\odot} \mathrm{yr}^{-1}$) just prior ($\sim$ yr -- decades) to their demise \citep[see ][and references therein]{2014ARA&A..52..487S}

Many interacting SNe have been detected, such as SNe IIn and SNe Ibn, which are considered to be powered by the interaction of the SN ejecta with the dense circumstellar material (CSM) \citep{1997ARA&A..35..309F, 2008MNRAS.389..113P, 2012Sci...337..927G, 2013A&A...555A..10T, 2017ApJ...835..140M}.
The pre-SNe mass loss rates estimated for these SNe are generally high \citep[$\dot{M} \gtrsim 10^{-4} M_{\odot} \mathrm{yr}^{-1}$;][]{2012ApJ...744...10K, 2014MNRAS.439.2917M}, which are much higher than the typical stellar wind for the red supergiants \citep{1988A&AS...72..259D, 2005A&A...438..273V}.
Some SNe are interpreted to experience the shock breakout within a dense CSM \citep{2010ApJ...724.1396O, 2013MNRAS.428.1020M}.
Moreover, for some SNe, the pre-SN stellar activities have been detected, probably related to the pre-SN mass loss, although the possibility remains that many of them  might not be the terminal explosion \citep{2007Natur.447..829P, 2013MNRAS.430.1801M, 2013Natur.494...65O, 2014ApJ...789..104O, 2017A&A...599A.129T}. 

The enhanced pre-SN mass loss may also be common for SNe II, which are defined to have hydrogen lines with the P-Cygni profile in their spectra. Recent high cadence surveys, such as the intermediate Palomar Transient Factory \citep{2009PASP..121.1395L}, have enabled us to catch SNe at the very early phase after the explosion. The spectra characterized by emission lines from highly-ionized ions (the so-called flash spectra) imply the elevated pre-SN mass loss for at least a fraction of SN IIP progenitors \citep{2017NatPh..13..510Y}. \citet{2016ApJ...818....3K} have found such flash-ionized spectra for $18 \%$ of their SNe II sample observed at ages $<$ 5 days, setting a lower limit for such phenomena.
Moreover, the early-time light curves of SNe II have been proposed to be better fit with dense CSM \citep{2017ApJ...838...28M, 2018NatAs.tmp..122F}. Thus, the enhanced pre-SN mass loss seems to be a common property.

The underlying mechanism of such an enhanced pre-SN mass loss is not well understood. 
It has been claimed that energy deposition into the envelope related to the advanced burning phases might be responsible for the enhanced mass loss \citep{2010MNRAS.405.2113D, 2014ARA&A..52..487S}. Various mechanisms have been proposed for the energy deposition. For example, a fraction of the gravity waves generated from the convective region in the core may tunnel towards the envelope, and deposit energy there \citep{2012MNRAS.423L..92Q, 2017MNRAS.470.1642F, 2018MNRAS.476.1853F}. The energy deposition rate by this process is expected to exceed Eddington luminosity only in the last few years before the core collapse. Thus, it can naturally explain the finely tuned timing of the event close to the core collapse \citep{2014ApJ...780...96S}. 
Explosive shell burning instabilities might also create additional energy \citep{2011ApJ...733...78A, 2014ApJ...785...82S}. Yet as another possibility, \citet{2012ApJ...752L...2C} has proposed that common envelope interaction as the cause of mass loss. In this hypothesis, a companion deposits its orbital energy into the primary's envelope, which then presumably unbinds the envelope.

However, many of these works mainly focus on demonstrating the validity of each idea, based on an order of magnitude estimate. How such an energy deposition affects the progenitor's density structure and SNe light curves has not been calculated consistently. \citet{2016MNRAS.458.1214Q} have investigated the effect of near-surface super-Eddington energy deposition on the structure of the envelope. They have shown that the extended wind are developed and the properties of the wind are consistent with analytic predictions. However, they considered only the energy deposition which takes place around the constant radius near the stellar surface. Moreover, the radiation-hydrodynamic calculation of SN based on the derived density structure of the progenitor has not been done. On the other hand, there have been a lot of studies which investigated the effect of CSM on the SNe light curves. They usually attach a density structure assuming a power law profile to the stellar surface, without considering how it is produced \citep{2011ApJ...729L...6C, 2011MNRAS.415..199M, 2017ApJ...838...28M}.


In this paper, we simulate the response of the envelope to various kinds of sustained energy deposition which hypothetically occurs within a few years before the core collapse, and investigate its effect on the nature of the SN progenitor. Furthermore, using the density profile thus derived, we calculate the light curves of the SNe self-consistently. 
In the present study, we do not specify those energy injection rates based on certain physical mechanisms. Rather, we artificially inject energy with parameterized forms, and investigate its effect in general. From these calculations, we aim to clarifying to what extent such energy deposition can explain the confined CSM for massive stars. Another purpose of this study is to constrain the nature of the pre-SN activity, by the requirement that the resulting progenitor and SN emission should be consistent with the existing data set, irrespective of the nature of the confined CSM. 

This paper is organized as follows: in \S \ref{method}, we describe the procedures of calculations, both for the stellar evolution using MESA (Modules for Experiments in Stellar Astrophysics) and for the radiation hydrodynamic simulation of the SNe using SNEC (SuperNova Explosion Code). In \S \ref{result}, we show the results of our calculations. Firstly, we focus on one model and discuss the effect of the energy deposition in general (\S \ref{rho_strucure_Ldep5d39uni} -- \S \ref{sec_vph_Ldep5d39}). Next, we investigate the effect of varying locations (\S \ref{sec_diff_locs}) and the rates (\S \ref{sec_different_rate}) of energy injection on the progenitors and SNe. We also discuss the effect of energy injection on the location of the progenitor on the HR diagram (\S \ref{sec_HR}).
In \S \ref{discussion}, we discuss the possible application of our results to peculiar SNe. We also discuss our results in relation to the hypothesis of gravity waves. Finally in \S \ref{conclusion}, we summarize the content of this paper.

\section{Method} \label{method}

\subsection{Hydrodynamic stellar evolution with energy injection using MESA}

\begin{table*}[htbp]
  \caption{Summary of progenitor and explosion properties of each of the different models.} \label{tab:decimal}
\hspace{-2.5cm}
  \scalebox{0.9}{
\begin{tabular}{lccccccccc}
\tablewidth{0pt}
\hline
Model     \footnote{Model name.}       &  log$ T_{\mathrm{eff}}$ \footnote{Effective temperature of the progenitor at the time of core collapse.} & log $L $ \footnote{Photospheric luminosity of the progenitor at the time of core collapse.} & $R_{\mathrm{ph}}$ \footnote{Photospheric radius of the progenitor at the time of core collapse.} & $R_{\mathrm{out}}$ \footnote{The radial coordinate of the outermost numerical cell of the progenitor at the time of core collapse.} &  $E_{\mathrm{bin, env}}$ \footnote{Binding energy of the envelope of the progenitor at the time of core collapse. This is calculated as $E_{\mathrm{bin, env}} = \int_{M_{\mathrm{He}}}^{M_{\mathrm{total}}} Gm/r \ dm$, where $M_{\mathrm{total}}$ and $M_{\mathrm{He}}$ are the total mass and He core mass of the progenitor, respectively. } & Unbound mass \footnote{The integrated mass of the progenitor at the time of core collapse for the region in the envelope where the specific total energy is positive. Here, the total energy is the sum of the kinetic energy, internal energy, and the gravitational potential energy.} & $t_{\mathrm{SB}}$ \footnote{Time since the explosion to the shock breakout.} &   $L_{\mathrm{pt}, 50}$   \footnote{Luminosity of the supernova at 50 days since the shock breakout.} & Plateau duration  \footnote{Duration of plateau. This is calculated as the time interval between the time of shock breakout and the time when the luminosity becomes half of $L_{\mathrm{pt}, 50}$.} \\
& (K)                      &  ($L_{\odot}$)           & ($R_{\odot}$) & ($R_{\odot}$) &  ($10^{47}$ erg) & ($M_{\odot}$) & (day) &  ($10^{42}$ erg s$^{-1}$) & (day) \\
\hline
\hline
No injection    &  3.52             & 4.93              &  876.3     &   926.9       &  8.67     &  0.00       & 1.68   &  2.35    & 89.1   \\
Ldep5d39uni      & 3.53           & 6.18              &  3553.2   &  5571.0        &  2.14     &  7.14       & 9.02   & 9.04     & 98.7 \\
Ldep5d39base     &  3.54             & 6.15        & 3282.7  & 3658.4       &  0.91     &  7.36       & 5.49  & 7.23    & 114.6 \\
Ldep5d39middle   &  3.54             &  6.17             &  3363.2   &  6081.2     &  3.10     &  6.30       & 9.07   & 8.96     & 113.6 \\
Ldep1d38uni      &  3.52             &  4.96             &  906.4    &  960.6         & 8.43      &  0.00       & 1.75   &  2.34    & 87.9 \\
Ldep1d39uni      &  3.51             &  5.23             &  1316.7   & 1434.7       & 5.94      &  0.00       & 2.56   & 3.58     & 84.6  \\
Ldep1d40uni      &  3.59  &  5.89             &  1899.7   & 8977.9          & 1.25      &  6.73       & 12.88  & 12.33    & 105.0  \\
\hline
\end{tabular}
}
\label{table}
\end{table*}

\begin{figure}[htbp] 
     \hspace{-3.3cm}
     \includegraphics[width=16cm, bb = 50 50 410 297]{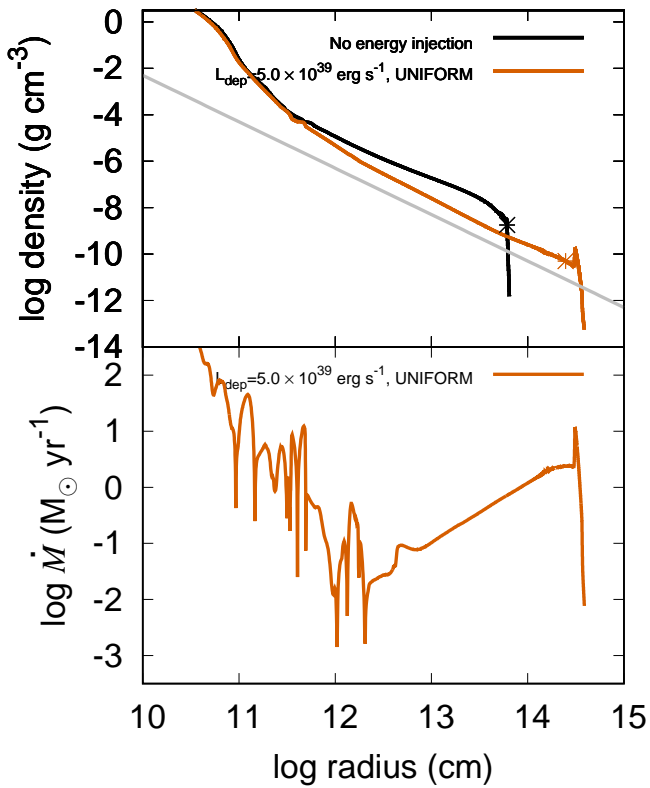}
\caption{Top: the density profile at the time of core collapse. The model with $L_{\mathrm{dep}}= 5.0 \times 10^{39}$ erg s$^{-1}$ and UNIFORM distribution is shown with a red solid line, while a black line shows the density profile of the model without the pre-SN energy injection. For reference, the density profile for a constant mass flux of $\dot{M} = 10^{-1} M_{\odot} \mathrm{yr}^{-1}$ is also shown with a gray line, assuming the constant wind velocity of $v = 10$ km s$^{-1}$. The asterisk marks show the locations of the photosphere, where $\tau = 2/3$, for each model. Bottom: the mass flux $\dot{M}=4 \pi r^2 \rho v$ for the model with $L_{\mathrm{dep}}= 5.0 \times 10^{39}$ erg s$^{-1}$ and UNIFORM distribution (red).}
  \label{density_Ldep5d39}
\end{figure}

\begin{figure}[t] 
  \begin{center}
  \includegraphics[width=9cm, bb = 50 50 410 300]{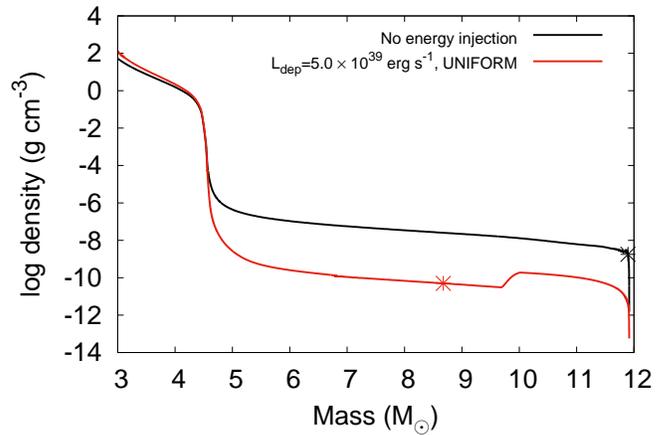}
  \caption{The density profile at the time of core collapse plotted as a function of mass coordinate. The model with $L_{\mathrm{dep}}= 5.0 \times 10^{39}$ erg s$^{-1}$ and UNIFORM distribution is shown with a red solid line, while a black line shows the density profile of the model without the pre-SN energy injection. The asterisk marks show the locations of the photosphere, where $\tau = 2/3$, for each model.
  }
  \label{mass_vs_rho}
  \end{center}
\end{figure}

\begin{figure*}[hbtp]
 \begin{center}
    \begin{tabular}{c}
    \begin{minipage}{0.5\hsize}
    \begin{center}
      \includegraphics[width=90mm, bb = 50 50 410 300]{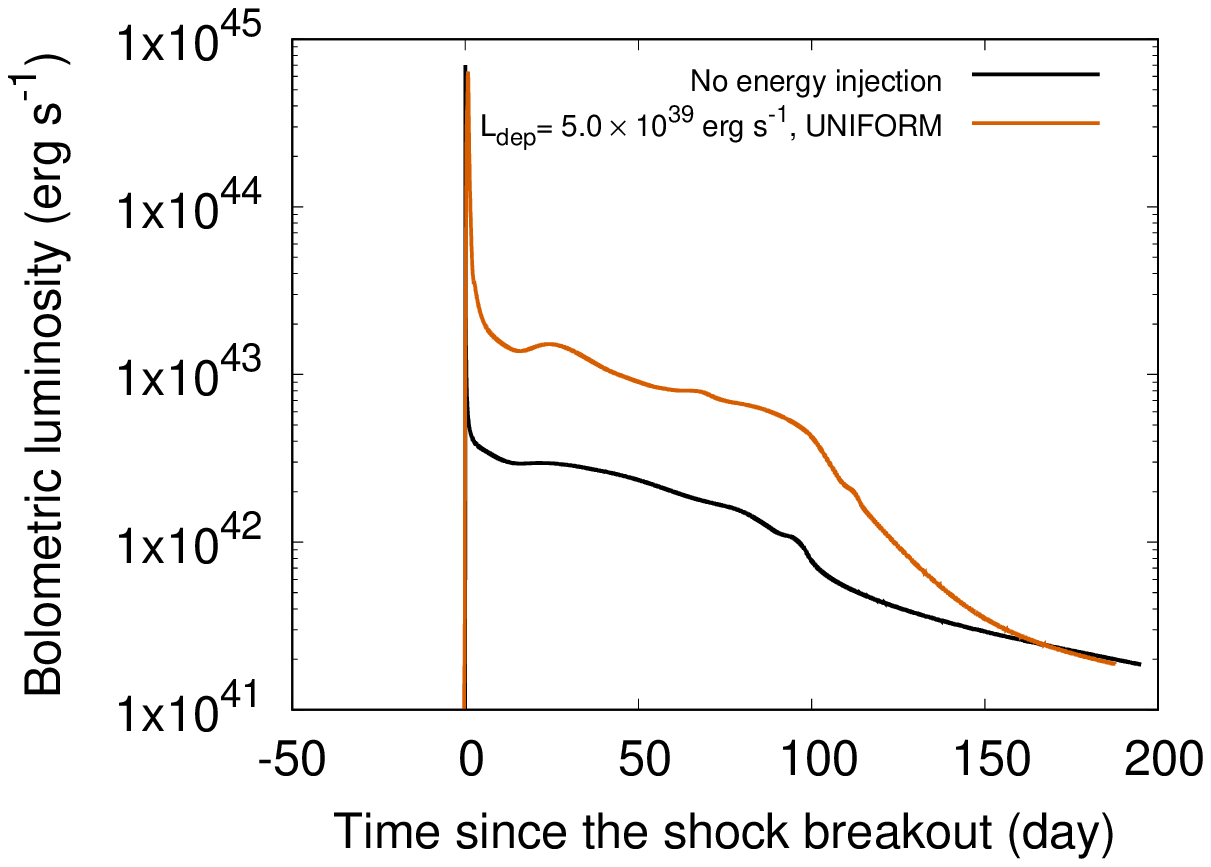}
    \end{center}
  \end{minipage}
  \begin{minipage}{0.5\hsize}
    \begin{center}
       \includegraphics[width=90mm, bb = 50 50 410 300]{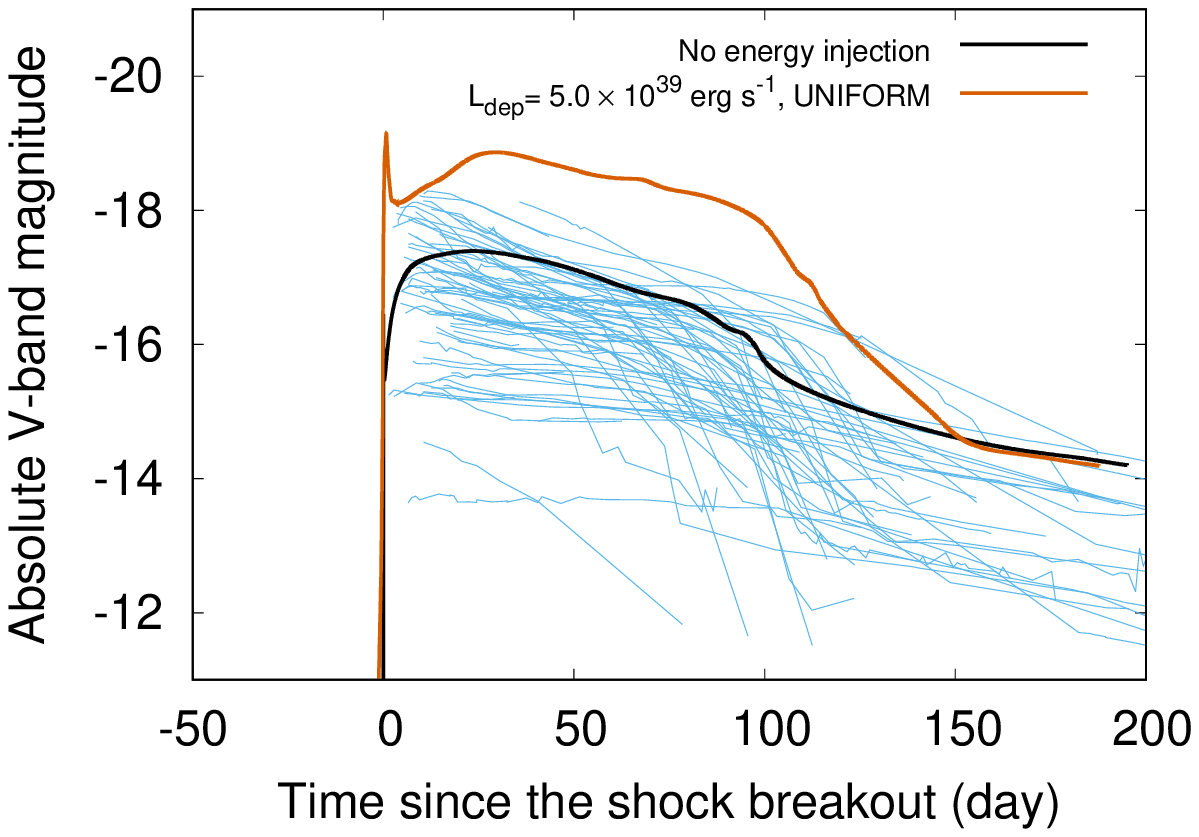}
    \end{center}
  \end{minipage}
  \end{tabular}
 \end{center}
 \caption{Left: the bolometric light curve of the model with $L_{\mathrm{dep}}= 5.0 \times 10^{39}$ erg s$^{-1}$ and UNIFORM distribution (red). A black line shows the model without the pre-SN energy injection. Right: the V-band absolute magnitude of the same models with the same colors as the left panel. The observational data for SNe II are shown with blue lines, which are taken from \citet{2014ApJ...786...67A}. The zero point of the $x$-axis for the models corresponds to the time of shock breakout, while for the observational data, it corresponds to the time of the estimated explosion epoch.}
  \label{LC_Ldep5d39uni}
\end{figure*}

For the calculation of stellar evolution, we use the one-dimensional stellar evolution code MESA of version 10398 \citep{2011ApJS..192....3P, 2013ApJS..208....4P, 2015ApJS..220...15P, 2018ApJS..234...34P}. We assume the initial metalicity of Z=0.02. 

First, we evolve a $15M_{\odot}$ non-rotating star from pre-main sequence to 3.0 years before the core collapse, without energy injection. Then, we start injecting energy into the envelope, with the hydrodynamic mode of MESA on, and evolve the model until the time of core collapse. The timescale of 3.0 years is chosen to represent the timescale of the pre-SN activities suggested by the observational studies \citep{2017NatPh..13..510Y}. This timescale is also consistent with the timescale for which the energy injection in the envelope by gravity waves is predicted to be significant \citep{2014ApJ...780...96S, 2017MNRAS.470.1642F}. The detailed settings for the calculation are explained in Appendix \ref{appendix_A}.

For the spatial distribution of the energy injection rate per mass $\epsilon_{\mathrm{inject}}$, we consider three cases: the uniform deposition (UNIFORM), the deposition into the base of envelope (BASE) and the deposition with a gaussian distribution (MIDDLE). In the case of UNIFORM, we uniformly inject energy at an injection rate of $L_{\mathrm{dep}}$ into the envelope, from just above the He core up to the stellar surface. At the inner boundary of the injected region, we linearly decline the energy injection rate to zero with the width of $0.01M_{\odot}$. In the case of BASE, we uniformly inject energy at a rate of $L_{\mathrm{dep}}$ into the region from just above the He core up to $0.1M_{\odot}$ above it. At both the outer and inner boundaries of the injected region, we linearly decline the energy injection rate to zero with the width of $0.01M_{\odot}$. Lastly, in the case of MIDDLE, we assume  the gaussian function for the energy injection rate per mass:
\begin{equation}
\epsilon_{\mathrm{inject}} = \frac{L_{\mathrm{dep}}}{\sigma \sqrt{2 \pi}} \exp \left(-\frac{(r - R_{\mathrm{dep}})^2}{2 \sigma^2} \right) \frac{dr}{dm},
\end{equation}
where $dr$, $dm$ are the width of a numerical cell in the radial and mass coordinates, respectively. We assume $R_{\mathrm{dep}}=500 R_{\odot}$, and $\sigma = 10.0 R_{\odot} $. The stellar radius of the progenitor before the energy injection is $\sim 880R_{\odot}$. So, this case represents a situation in which the energy deposition takes place roughly in the middle of the envelope.

For the energy injection rate $L_{\mathrm{dep}}$, we investigate four values: $L_{\mathrm{dep}} = 1.0 \times 10^{38} $ erg s$^{-1}$ $, 1.0 \times 10^{39} $ erg s$^{-1}$ $, 5.0 \times 10^{39} $ erg s$^{-1}$, and $1.0 \times 10^{40} $ erg s$^{-1}$. These values are chosen to be around the Eddington luminosity of the progenitor. 
\begin{figure*}[htbp] 
 \begin{center}
    \begin{tabular}{c}
    \begin{minipage}{0.5\hsize}
    \begin{center}
       \includegraphics[width=90mm, bb =50 50 410 300]{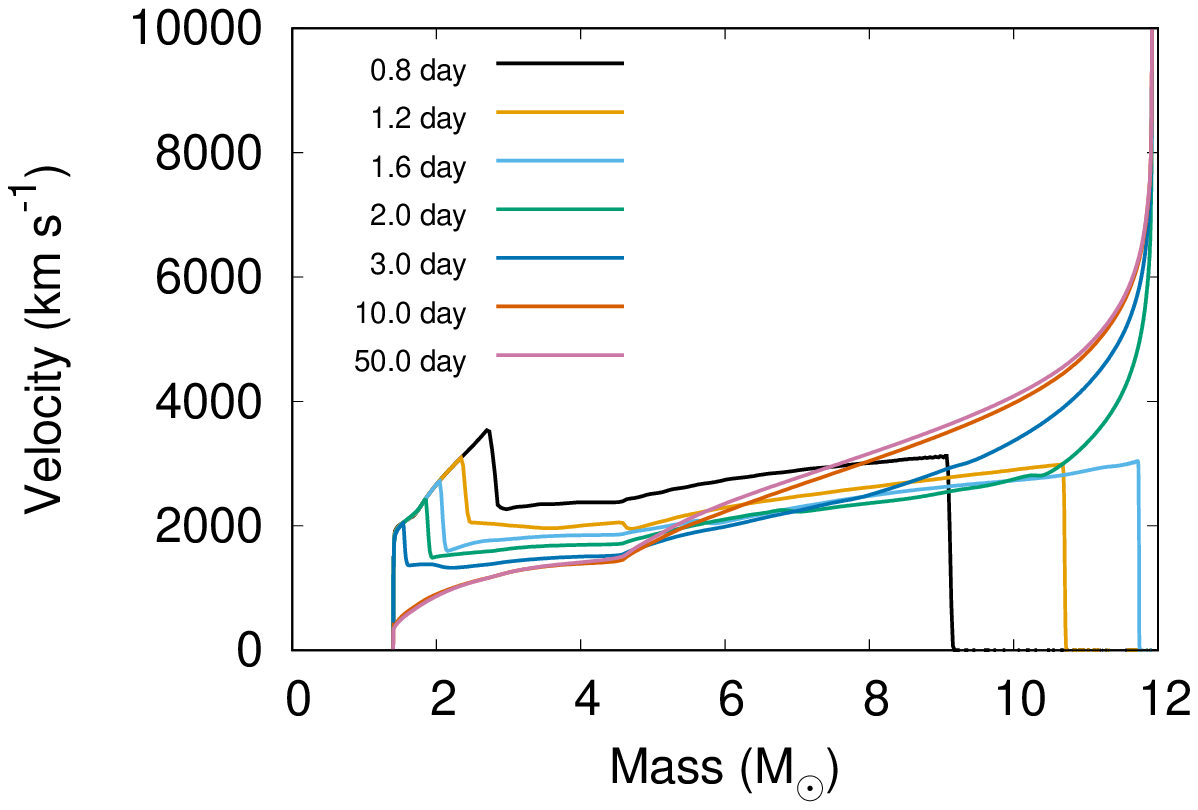}
    \end{center}
  \end{minipage}
  \begin{minipage}{0.5\hsize}
    \begin{center}
       \includegraphics[width=90mm, bb =50 50 410 300]{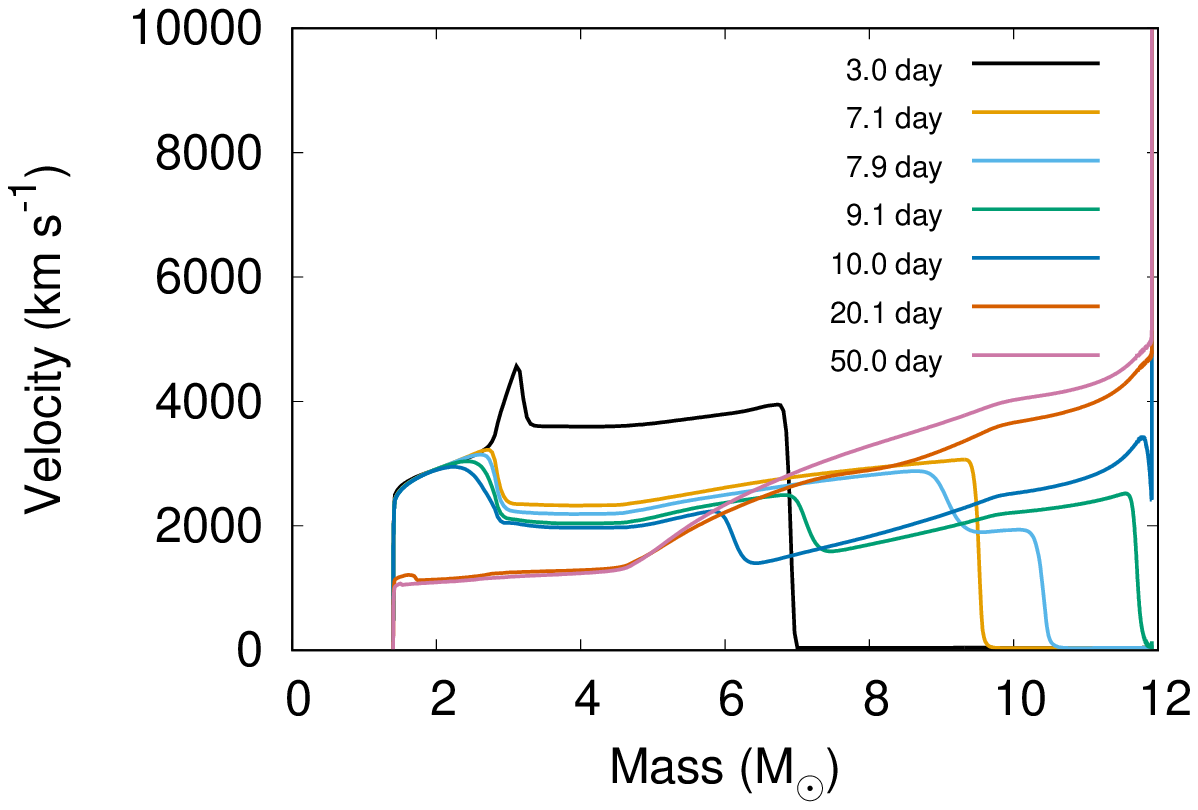}
    \end{center}
  \end{minipage}
  \end{tabular}
 \end{center}
 \vspace{-0.3cm}
 \caption{Left: The time evolution of the velocity profile for the model without the pre-SN energy injection. Right: The same plot as the left panel for the model with $L_{\mathrm{dep}} = 5.0 \times 10^{39} $ erg s$^{-1}$ and UNIFORM distribution. For both panels, the zero point of time corresponds to the time of explosion.}
  \label{vel_evolution}
\end{figure*}

\begin{figure*}[htbp] 
    \begin{tabular}{c}
    \begin{minipage}{0.5\hsize}
    \begin{center}
      \includegraphics[ width=90mm, bb =50 50 410 300]{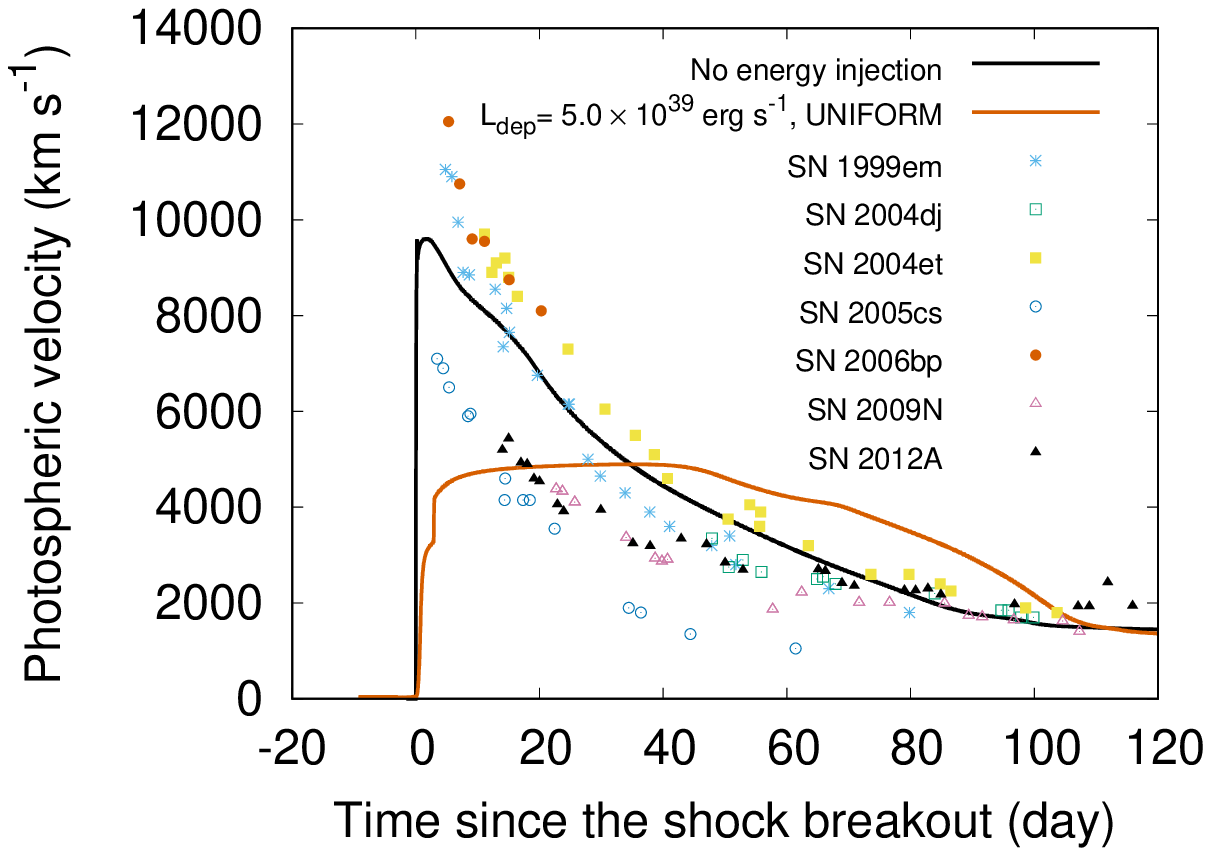}
    \end{center}
  \end{minipage}
  \begin{minipage}{0.5\hsize}
    \begin{center}
       \includegraphics[width=90mm, bb =50 50 410 300]{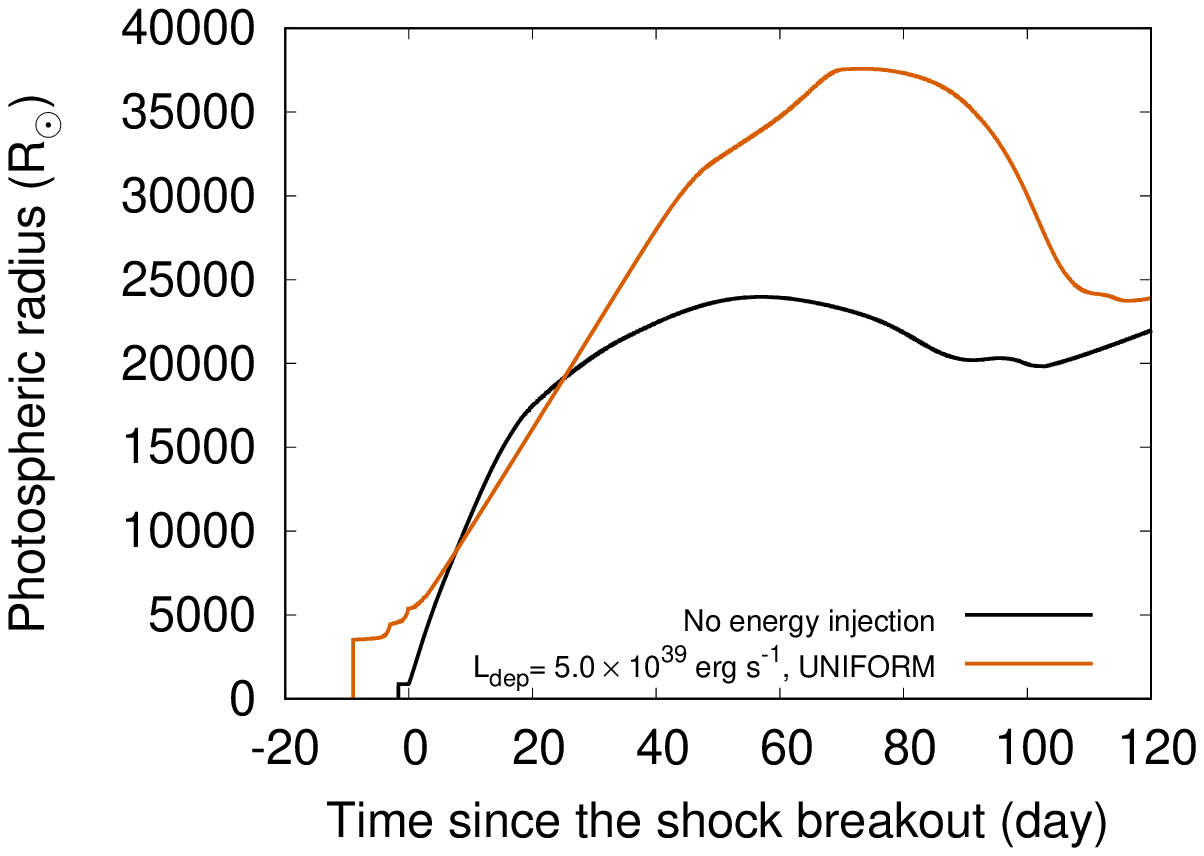}
    \end{center}
  \end{minipage}
  \end{tabular}
 \vspace{0.1cm}
 \caption{Left: The photospheric velocity of the model with $L_{\mathrm{dep}}= 5.0 \times 10^{39}$ erg s$^{-1}$ and UNIFORM distribution (red). A black line shows the model without the pre-SN energy injection. The observationally derived value for several SNe II are also plotted. The data for SN1999em, SN2004dj, SN2004et, SN2005cs, and SN2006bp are taken from \citet{2012MNRAS.419.2783T}. The data for SN2009N and SN2012A are taken from \citet{2014MNRAS.438..368T} and \citet{2013MNRAS.434.1636T}, respectively. The zero point of the $x$-axis for the models corresponds to the time of shock breakout, while for the observational data, it corresponds to the time of the estimated explosion epoch. Right: The time evolution of the photospheric radius of the same models as the left panel with the same colors.}
  \label{vph_and_Rph_Ldep5d39}
\end{figure*}

\subsection{Radiation hydrodynamic simulation of SNe using SNEC}
Once the models described above are evolved dynamically to the time of the core-collapse, they are used as input models for the radiation hydrodynamic simulations of the SN explosions. For this purpose, we use the open code SNEC \citep{2015ApJ...814...63M} \footnote{http://stellarcollapse.org/SNEC}. SNEC is a one-dimensional Lagrangian hydrodynamic code, which also solves radiation energy transport with the flux-limited diffusion approximation. The code generates the bolometric light curves of the SNe, as well as the light curves in different observed wavelength bands in the blackbody approximation. The number of the cells is set to be 1000.

First, we excise the innermost $1.4 M_{\odot}$ before the explosion, assuming that it collapses to form a neutron star.
Then, we simulate the explosion as a thermal bomb, by adding the energy in the inner $0.1 M_{\odot}$ of the model for a duration of 0.1s. For the explosion energy, we fix it to $1.0 \times 10^{51} \mathrm{erg}$, a typical value for SNe II \citep{2003ApJ...582..905H, 2011ApJ...729...61B, 2011MNRAS.417..261I, 2013A&A...555A.142I}.

The compositional profiles are smoothed using the “boxcar” approach. This mimics the mixing due to the Rayleigh-Taylor instabilities during the explosion. We run a boxcar with a width of $0.4 M_{\odot}$ through the model four times until a smooth profile is obtained. The Rosseland mean opacity for each grid is taken from the existing tables \citep{1996ApJ...464..943I, 2005ApJ...623..585F}, taking the opacity floor into account. Following \citet{2015ApJ...814...63M}, the opacity floor is set to be linearly proportional to metalicity $Z$ at each grid point, and set it to 0.01 $\mathrm{cm}^2 \mathrm{g}^{-1}$ for solar composition ($Z_{\odot}=0.02$) and to 0.24 $\mathrm{cm}^2 \mathrm{g}^{-1}$ for a pure metal composition ($Z = 1$). We employ the analytic equation of state given by \citet{1983ApJ...267..315P}, which contains contributions from radiation, ions, and electrons. We trace the ionization fractions of hydrogen and helium solving the Saha equations. 

The code does not include a nuclear reaction network, and $^{56}$Ni is given by hand. Here, the mass of Ni is assumed to be $M_{\mathrm{Ni}}=0.07M_{\odot}$, which is a characteristic value for observed SNe II \citep{2003ApJ...582..905H, 2009ARA&A..47...63S, 2014ApJ...787..139D, 2017ApJ...841..127M}. Then, it is distributed from the inner boundary up to the mass coordinate of $m (r) =3M_{\odot}$.



\section{Result} \label{result}

In Table \ref{table}, we summarize the progenitor and explosion properties for each of the different models. Here, the model name of Ldep5d39middle, for example, denotes the model with $L_{\mathrm{dep}} = 5.0 \times 10^{39}$ erg s$^{-1}$, and MIDDLE distribution. For all the models, the He core mass and total mass of the progenitors at the time of core collapse are $4.56M_{\odot}$ and $11.9M_{\odot}$, respectively.

We first focus on the model with the energy injection rate of $L_{\mathrm{dep}}= 5.0 \times 10^{39} $ erg s$^{-1}$ with UNIFORM distribution (hereafter, Ldep5d39uni). This is to demonstrate how the sustained energy deposition in the envelope, continuing for a few years before the core collapse, affects the density profile of the progenitors and SNe.

\subsection{Density profile of the progenitor at the time of core collapse} \label{rho_strucure_Ldep5d39uni}

\begin{figure}[htbp]
  \includegraphics[width=9cm, bb =50 50 410 310]{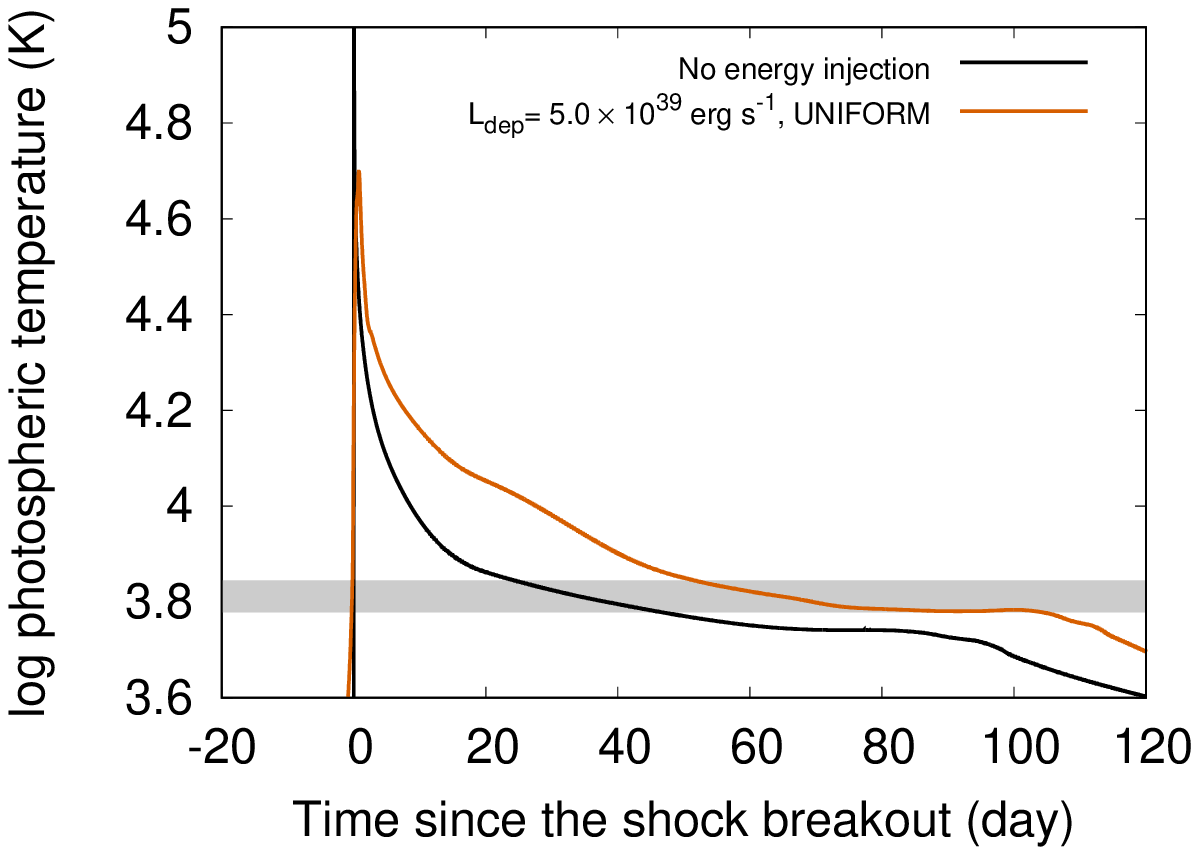}
  \caption{
      The time evolution of the photospheric temperature of the model with $L_{\mathrm{dep}}= 5.0 \times 10^{39}$ erg s$^{-1}$ and UNIFORM distribution (red). A black line shows the model without the pre-SN energy injection. Hatched area shows the region where the temperature is between $6000$ K and $7000$ K, which corresponds to the hydrogen recombination temperature.
  }
  \label{Tph_Ldep5d39}
\end{figure}

\begin{figure*}[hbtp]
    \begin{tabular}{c}
    \begin{minipage}{0.5\hsize}
    \begin{center}
      \includegraphics[ width=90mm, bb =50 50 410 300]{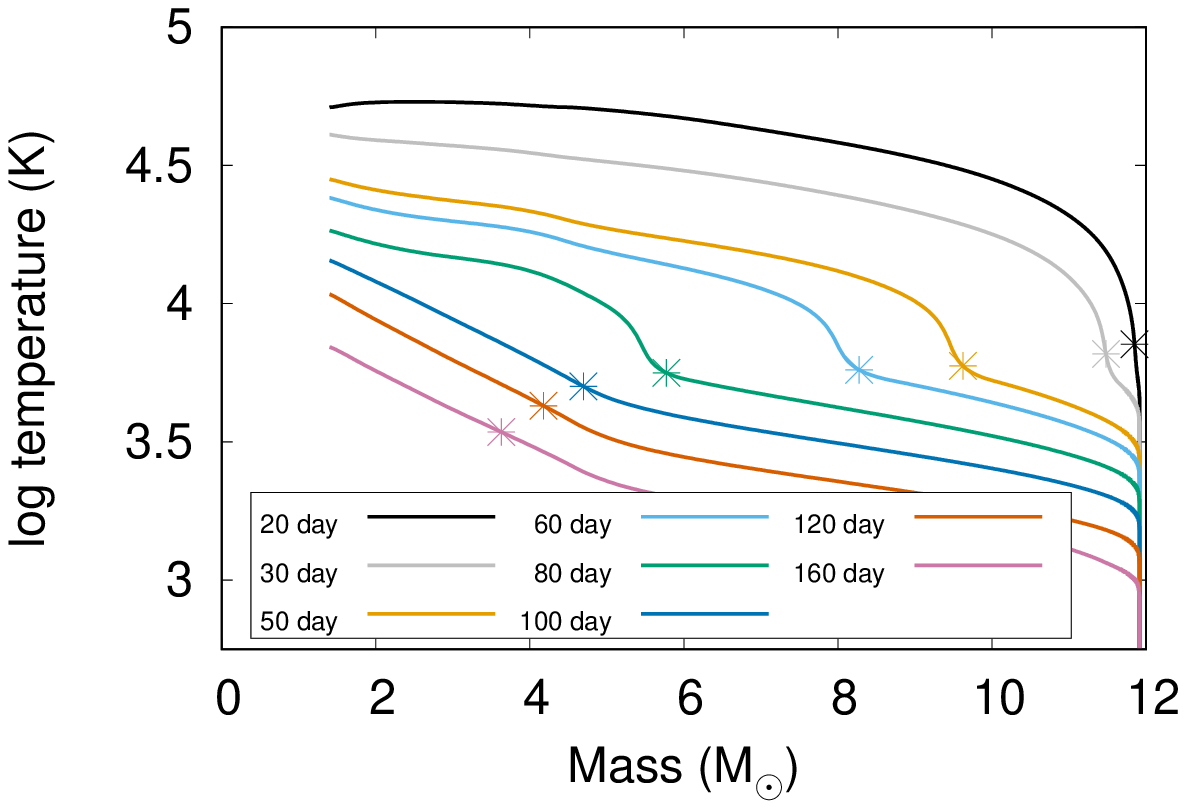}
    \end{center}
  \end{minipage}
  \begin{minipage}{0.5\hsize}
    \begin{center}
       \includegraphics[width=90mm, bb =50 50 410 300]{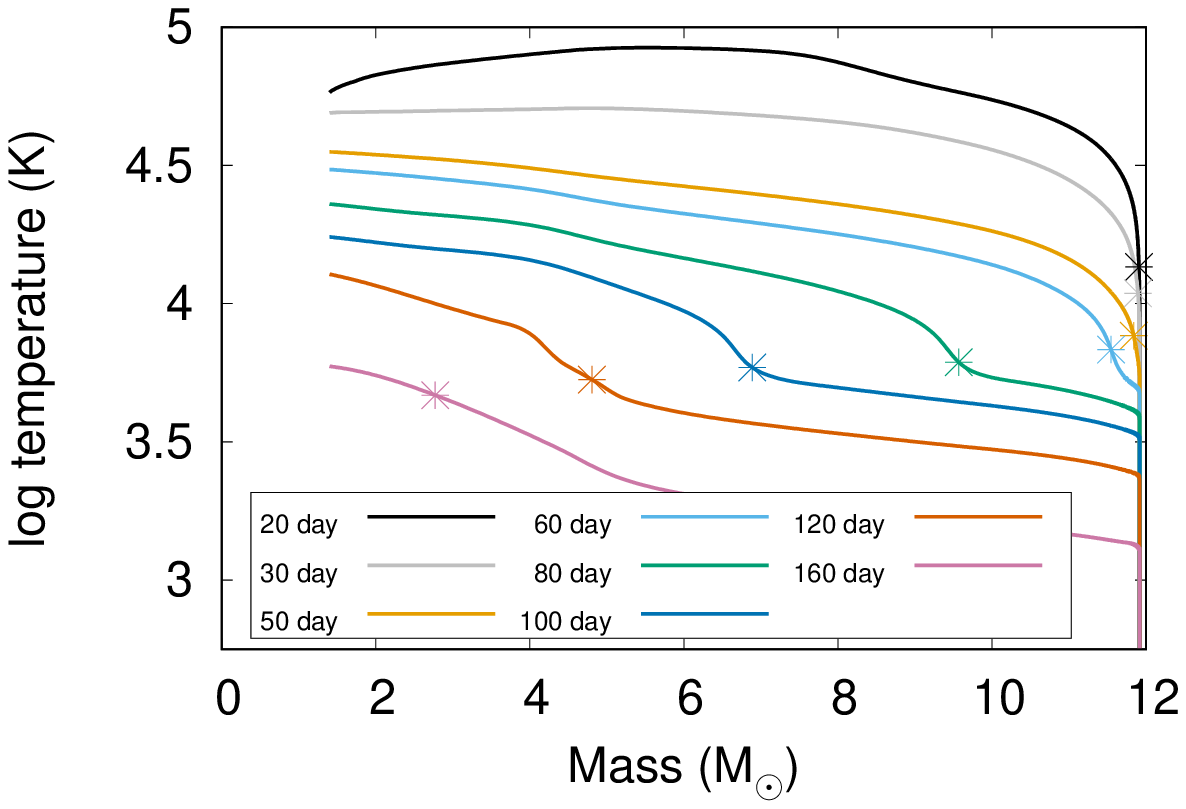}
    \end{center}
  \end{minipage}
  \end{tabular}
 \caption{Left: The time evolution of the temperature profile for the model without the pre-SN energy injection. Right: The same plot as the left panel for the model with $L_{\mathrm{dep}} = 5.0 \times 10^{39} $ erg s$^{-1}$ and UNIFORM distribution. For both panels, the zero point of time corresponds to the time of explosion, and the locations of the photosphere, where $\tau = 2/3$, are denoted by asterisk marks.}
  \label{T_evolution}
\end{figure*}

Fig.\ref{density_Ldep5d39} shows the density profile at the time of core collapse of the model, Ldep5d39uni, plotted as a function of radius. 
For reference, the density profile for a constant mass flux of $\dot{M} = 10^{-1} M_{\odot} \mathrm{yr}^{-1}$, assuming the constant wind velocity of $v = 10$ km s$^{-1}$, is also shown.
The envelope expands significantly, by converting the added heat to the pressure work, with the resulting mass flux of $\dot{M} \gtrsim 1 M_{\odot} \mathrm{yr}^{-1}$ above the initial stellar surface. The established wind above the initial surface is close to a steady state in this case, thus the density is nearly following $\rho \propto r^{-2}$. However, the wind-like CSM structure which is separated by the density discontinuity from the stellar surface is not produced.

Fig.\ref{mass_vs_rho} shows the density profile for the same model, plotted as a function of mass coordinate. There is a high density shell structure at the outer part of the envelope, which contains $\sim 2 M_{\odot}$. The wind velocity increases as time passes due to the lowered density, and the matter launched later catches up with the previously ejected matter, thus making a dense shell. This qualitative behaviour is consistent with the result of \citet{2016MNRAS.458.1214Q}, who have investigated the response of the envelope to the near-surface energy deposition using MESA.

The photosphere is located close to the outer edge of the wind (see the asterisk mark in Fig.\ref{density_Ldep5d39}), and the density structure is that of ``expanded envelope''. Actually, the photospheric radius of the model Ldep5d39uni is $3553.2 R_{\odot}$, which is much larger than the typical red supergiants. This qualitative result is consistent with the previous work by \citet{2014MNRAS.445.2492M}, who also concluded that the wave energy deposition in the envelope causes envelope expansion rather than mass ejection. Also, as is shown in the Table\ref{table}, the unbound mass of the progenitor for the model Ldep5d39uni is as much as $7.14M_{\odot}$, which is very close to the envelope mass of $7.36 M_{\odot}$. Thus, almost whole the envelope gets unbound due to the energy injection. Also, the inner radius of the unbound region is well within the photosphere for all the models that have unbound mass. For the model of Ldep5d39uni, for example, the inner radius of the unbound region is $171.9 R_{\odot}$, which is much smaller than the photospheric radius. Thus, the photosphere resides in the unbound region.


\subsection{The light curves} \label{sec_LC_Ldep5d39uni}

Fig.\ref{LC_Ldep5d39uni} compares the bolometric and $V$-band light curves of an SN for the model Ldep5d39uni to those for the model without the pre-SN energy injection. The left panel compares the bolometric light curve of each model. The model Ldep5d39uni reaches the shock breakout at $9.0$ days after the explosion, while it takes only $1.7$ days for the model without the pre-SN energy injection (see Table 1). Due to the envelope inflation, the shock has to travel larger distance in order to allow the shock energy to diffuse out. It can also be seen in Fig. \ref{vel_evolution}, which shows the time evolution of velocity profile for two models.

The plateau of the model Ldep5d39uni is brighter and longer than the model without the pre-SN energy injection. Both of these results can be explained by the longer expansion timescale for the model of Ldep5d39uni, due to the larger initial radius. Here, the expansion timescale is defined as $t_e \equiv R_0/v_{\mathrm{SN}}$, where $R_0$ is the initial progenitor radius and $v_{\mathrm{SN}}$ is the velocity of the SN ejecta. Because of the longer expansion timescale, the adiabatic cooling of the envelope is slower \citep{2009ApJ...703.2205K}. Therefore, the envelope retains more internal energy at the same epoch since the shock breakout and the SN has higher luminosity. Also, the slow cooling delays the propagation of the hydrogen recombination front, thus, prolonging the plateau phase.


The right panel of Fig.\ref{LC_Ldep5d39uni} compares the V-band absolute magnitude of the same SN models as the left panel. Here, the observational data of Type II SNe are also plotted, which are taken from \citet{2014ApJ...786...67A}. It is clear that the model Ldep5d39uni can not explain the light curves of Type II SNe, because the plateau is too bright and too long.



\subsection{Photospheric velocity} \label{sec_vph_Ldep5d39}

The left panel of Fig.\ref{vph_and_Rph_Ldep5d39} compares the photospheric velocity of the model Ldep5d39uni to that of the model without the pre-SN energy injection. Also shown here are the observational data of several SNe IIP \citep{2012MNRAS.419.2783T, 2014MNRAS.438..368T, 2013MNRAS.434.1636T}

The model Ldep5d39uni has a relatively constant photospheric velocity during the first $\sim$ 50 days from the explosion. After that, it gradually declines. This behaviour clearly deviates from the observational data, just like the light curve does. The nearly constant photospheric velocity stems from the behaviour of the photosphere, which resides near the outer edge of the SN ejecta during this time.
The model Ldep5d39uni has longer expansion timescale and slower envelope cooling. Therefore, the hydrogen begins to recombine later. This is clearly seen from Fig. \ref{Tph_Ldep5d39}, which shows the time evolution of the photospheric temperature for the model Ldep5d39uni. Only after $t \sim 50$ day, the photospheric temperature goes down to the hydrogen recombination temperature of $\sim 6000$ -- $7000$ K and the envelope begins to recombine. This timing matches the time when the photospheric velocity begins to decline (Fig.\ref{vph_and_Rph_Ldep5d39}). The same behaviour is also seen in Fig.\ref{T_evolution}, which shows the time evolution of the temperature profile for the same model.

The velocity is lower ($\sim 5000$ km s$^{-1}$) during the first $\sim 50$ days than the model without the pre-SN energy injection. This is because the shock velocity is decelerated when the shock hits the high density shell, located at the outer part of the envelope. This can be seen by comparing the Fig.\ref{mass_vs_rho} and Fig.\ref{vel_evolution} for the model of Ldep5d39uni; when the shock reaches $m(r) \sim 10M_{\odot}$, where the high density shell is located, the shock is decelerated. Additionally, the shock has travelled farther radially outward, which also leads to lower velocity in the outermost ejecta \citep{1999ApJ...510..379M}.


The right panel of Fig.\ref{vph_and_Rph_Ldep5d39} compares the time evolution of the photospheric radius. Until $t \sim 50$ days, the radius keeps increasing linearly with time, and then the increase becomes slower after that. On the contrary, the photospheric radius of the model without the pre-SN energy injection keeps nearly constant after $t \gtrsim 20$ day. 

\subsection{Dependence on the location of energy injection} \label{sec_diff_locs}

So far, we have focused on the model Ldep5d39uni in order to clarify general effects of the energy injection on the properties of the SN and its progenitor. From now on, we compare different models to clarify how the outcomes are affected by the properties of the energy injection. In this section, we investigate how the location of energy injection affects the pre-SN density profile, the light curve and photospheric velocity of an SN. In this section, we fix $L_{\mathrm{dep}} = 5.0 \times 10^{39} $ erg s$^{-1}$, and vary the location of the energy injection in three ways as explained in \S \ref{method}: UNIFORM, BASE, and MIDDLE.

\subsubsection{Density profile of the progenitor at the time of core collapse}

\begin{figure}[t] 
  \hspace{-3.3cm}
  \includegraphics[width=16cm, bb =50 50 410 297]{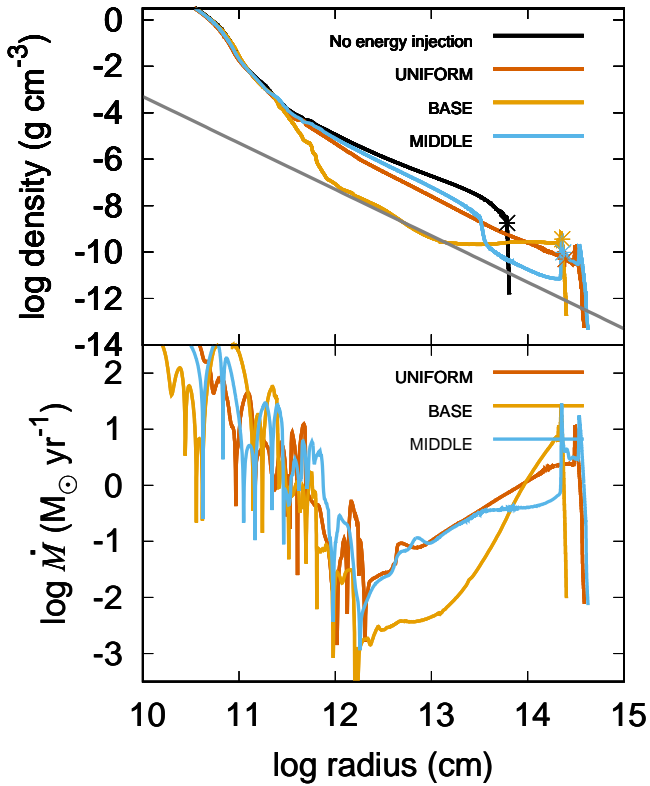}
  \caption{Top: the density profile at the time of core collapse of the models with different energy injection locations. A black line shows the model without the pre-SN energy injection. Red, orange, and blue solid lines denote the UNIFORM, BASE, and MIDDLE energy deposition, respectively. For reference, the density profile for a constant mass flux of $\dot{M} = 10^{-2} M_{\odot} \mathrm{yr}^{-1}$, assuming the constant wind velocity of $v = 10$ km s$^{-1}$ is also shown with a gray line. The asterisk marks denote the locations of the photosphere, where $\tau = 2/3$. Bottom: the mass flux for the three models, with the corresponding colors.}
  \label{density_profile_different_locations}
\end{figure}


\begin{figure*}[htbp]
    \begin{tabular}{c}
    \begin{minipage}{0.5\hsize}
    \begin{center}
       \includegraphics[width=90mm, bb =50 50 410 300]{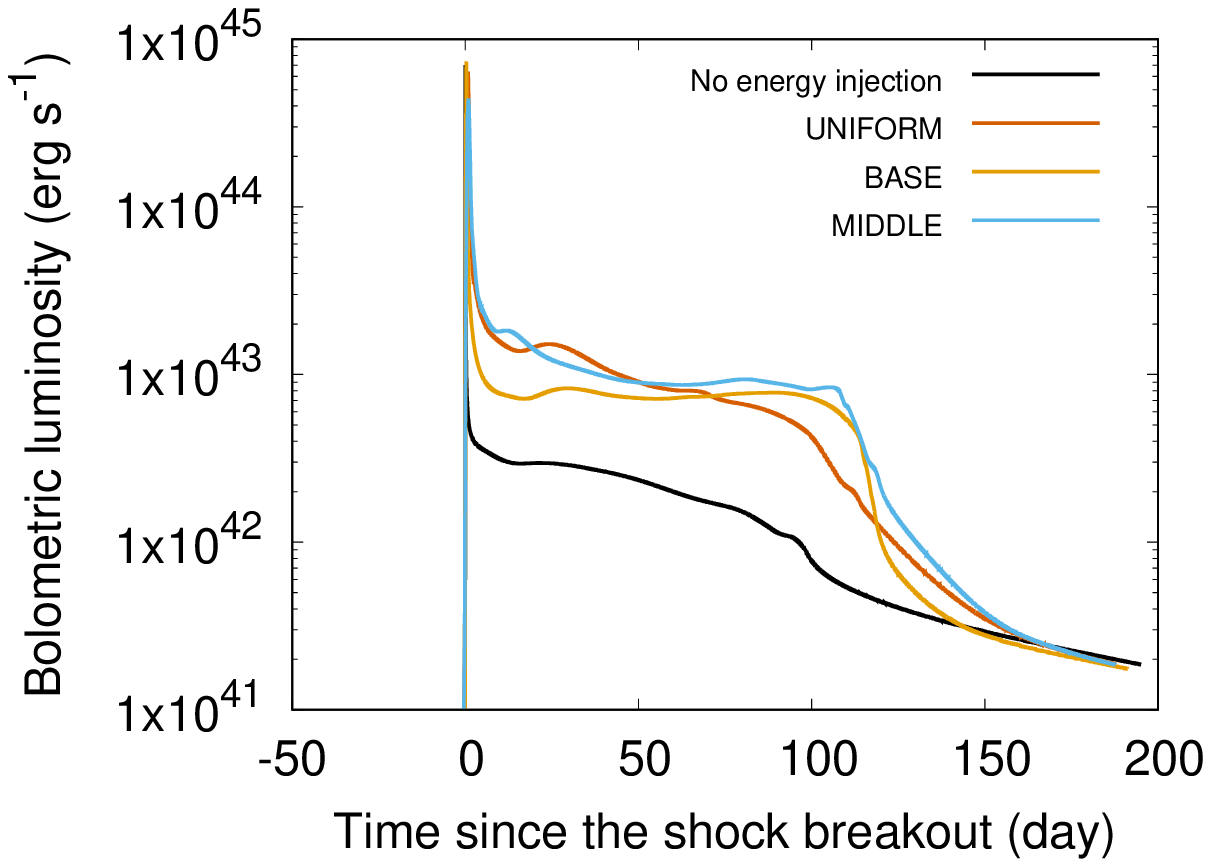}
    \end{center}
  \end{minipage}
  \begin{minipage}{0.5\hsize}
    \begin{center}
       \includegraphics[width=90mm, bb =50 50 410 300]{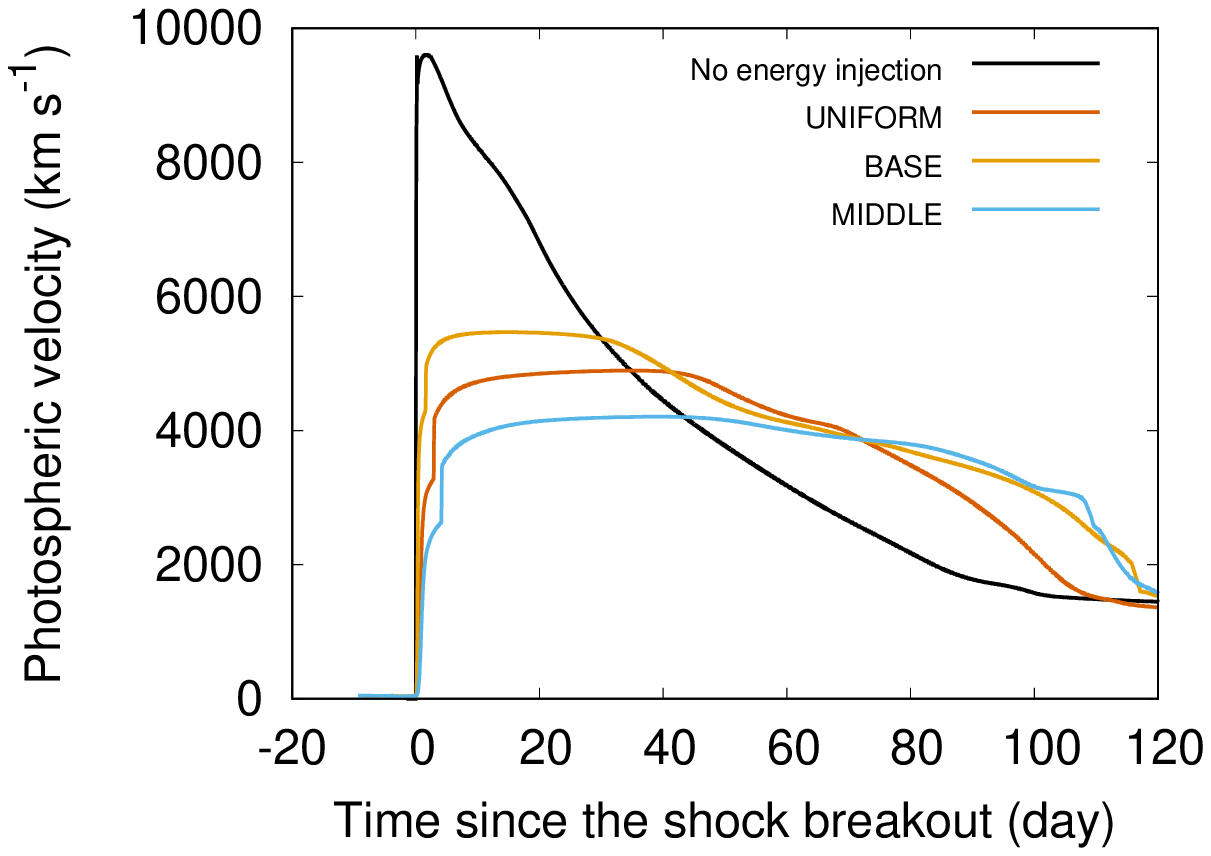}
    \end{center}
  \end{minipage}
  \end{tabular}
 \caption{Left: The bolometric light curve of the model with different locations for the energy injection. We fix $L_{\mathrm{dep}}= 5.0 \times 10^{39}$ erg s$^{-1}$. A black line shows the model without the pre-SN energy injection. Red, orange and blue lines show the model with UNIFORM, BASE, and MIDDLE energy deposition, respectively. Right: The photospheric velocity of the models, with the color, denoting likewise.}
  \label{LC_vph_different_location}
\end{figure*}

Fig.\ref{density_profile_different_locations} compares the density profile of the the models with different locations of the energy injection at the time of core collapse. 
For reference, the density profile for a constant mass flux of $\dot{M} = 10^{-2} M_{\odot} \mathrm{yr}^{-1}$, assuming the constant wind velocity of $v = 10$ km s$^{-1}$ is also shown.


In all the models with the pre-SN energy injection, the structure is composed of nearly steady wind structure, where $\rho \propto r^{-2}$, and a high density shell located above it. The reason for this structure was explained in \S \ref{rho_strucure_Ldep5d39uni}. The density structures are largely different between the three models. Especially for the case of BASE, the density structure of the envelope is quite different from the non-injecting case, with the resulting density lower by a few orders of magnitudes in the inner part of the envelope. Note, however, that the locations of the photosphere are nearly the same between the three models (Table\ref{table} and Fig.\ref{density_profile_different_locations}).


\subsubsection{Light curves and photospheric velocity}
The left panel of Fig.\ref{LC_vph_different_location} compares the bolometric light curves for the models with different locations of energy injection. All three models with pre-SN energy injection have brighter and longer plateau than the the model without the pre-SN energy injection. Overall shapes of the light curves are quite similar. In the case of UNIFORM and MIDDLE, a small bump soon after the major luminosity peak can be seen, while the effect is minor. In all the three cases, the outer part of the envelope are sufficiently dense, so that the shock breakout occurs only when the shock reaches the outer edge of the envelope. Thus, they behave like an expanded envelope.


The right panel of Fig. \ref{LC_vph_different_location} compares the photospheric velocity of the models with different locations of energy injection. Just like the light curves, the time evolution of photospheric velocity is similar irrespective of the location of the energy injection. All three models with the energy injection have relatively flat photospheric velocity evolution during the first few ten days from the explosion, which then gradually declines.

\subsection{Dependence on the energy injection rate} \label{sec_different_rate}
Next, we investigate how the different energy injection rates ($L_{\mathrm{dep}}$) affect the properties of the progenitors and resulting SNe. In this section, we assume the UNIFORM distribution for the energy injection rate.

\subsubsection{Density profile of the progenitor at the time of core collapse} \label{rho_different_rate}

\begin{figure}[htbp]
  \hspace{-3.3cm}
 \includegraphics[width=16cm, bb =50 50 410 297]{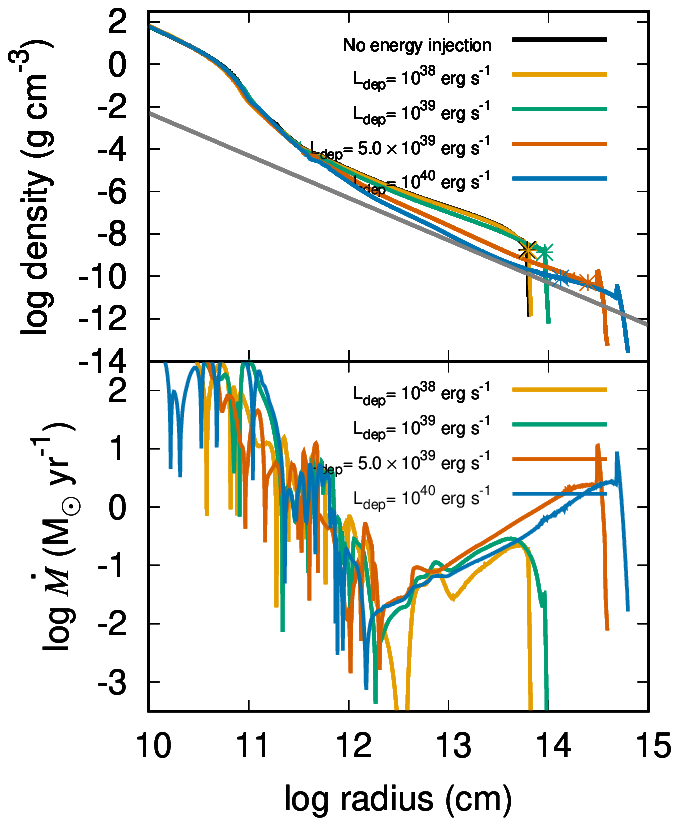}
 \caption{Top: the density profile at the time of core collapse of the models with different energy injection rates. A black line shows the model without the pre-SN energy injection. Orange, green, red, and blue lines show the model with $L_{\mathrm{dep}}= 10^{38} $ erg s$^{-1}$ $, 10^{39} $ erg s$^{-1}$ $, 5.0 \times 10^{39} $ erg s$^{-1}$ $, 10^{40} $ erg s$^{-1}$, respectively. For reference, the density profile for a constant mass flux of $\dot{M} = 10^{-1} M_{\odot} \mathrm{yr}^{-1}$, assuming the constant wind velocity of $v = 10$ km s$^{-1}$ is also shown with a gray line. The asterisk marks denote the locations of the photosphere, where $\tau = 2/3$. Bottom: the mass flux $\dot{M}=4 \pi r^2 \rho v$ for the models with $L_{\mathrm{dep}}= 10^{38} $ erg s$^{-1}$ $, 10^{39} $ erg s$^{-1}$ $, 5.0 \times 10^{39} $ erg s$^{-1}$ $, 10^{40} $ erg s$^{-1}$, with the same colors as above.}
  \label{density_profile_different_rate}
\end{figure}

\begin{figure*}[htbp]
    \begin{tabular}{c}
    \begin{minipage}{0.5\hsize}
    \begin{center}
       \includegraphics[width=90mm, bb =50 50 410 300]{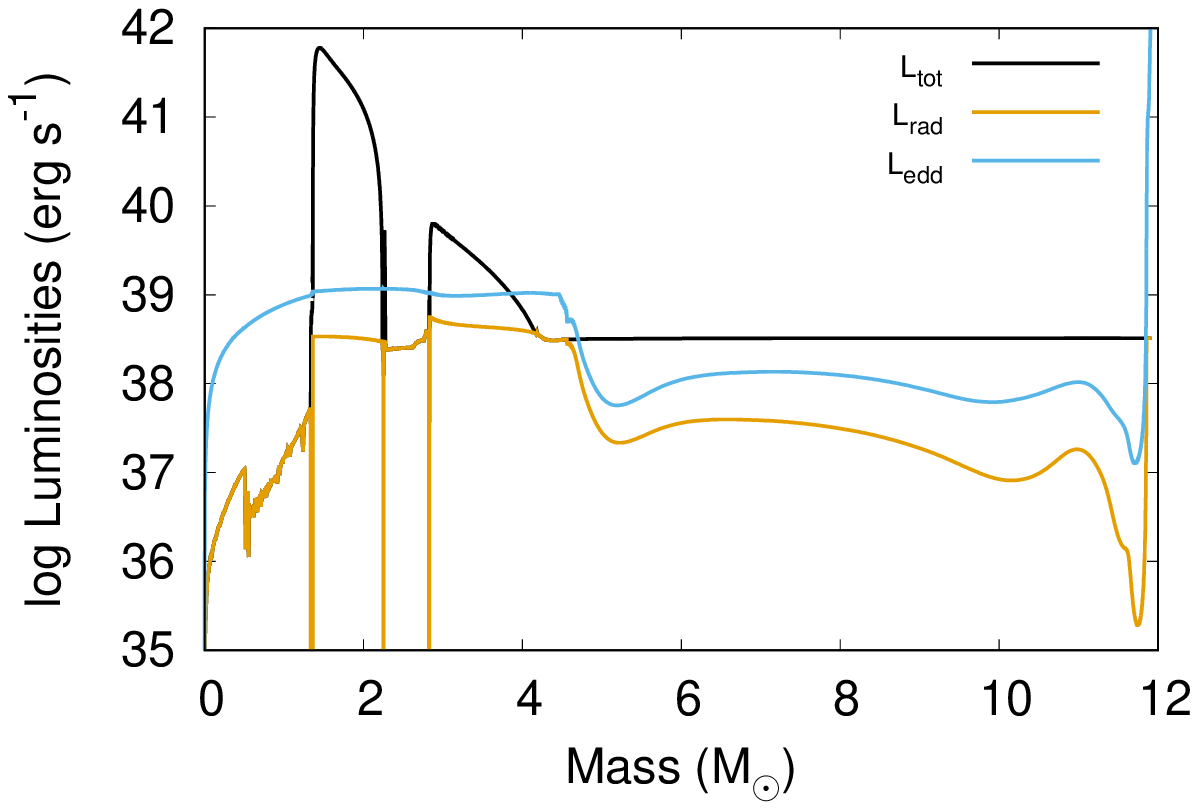}
    \end{center}
  \end{minipage}
  \begin{minipage}{0.5\hsize}
    \begin{center}
       \includegraphics[width=90mm, bb =50 50 410 300]{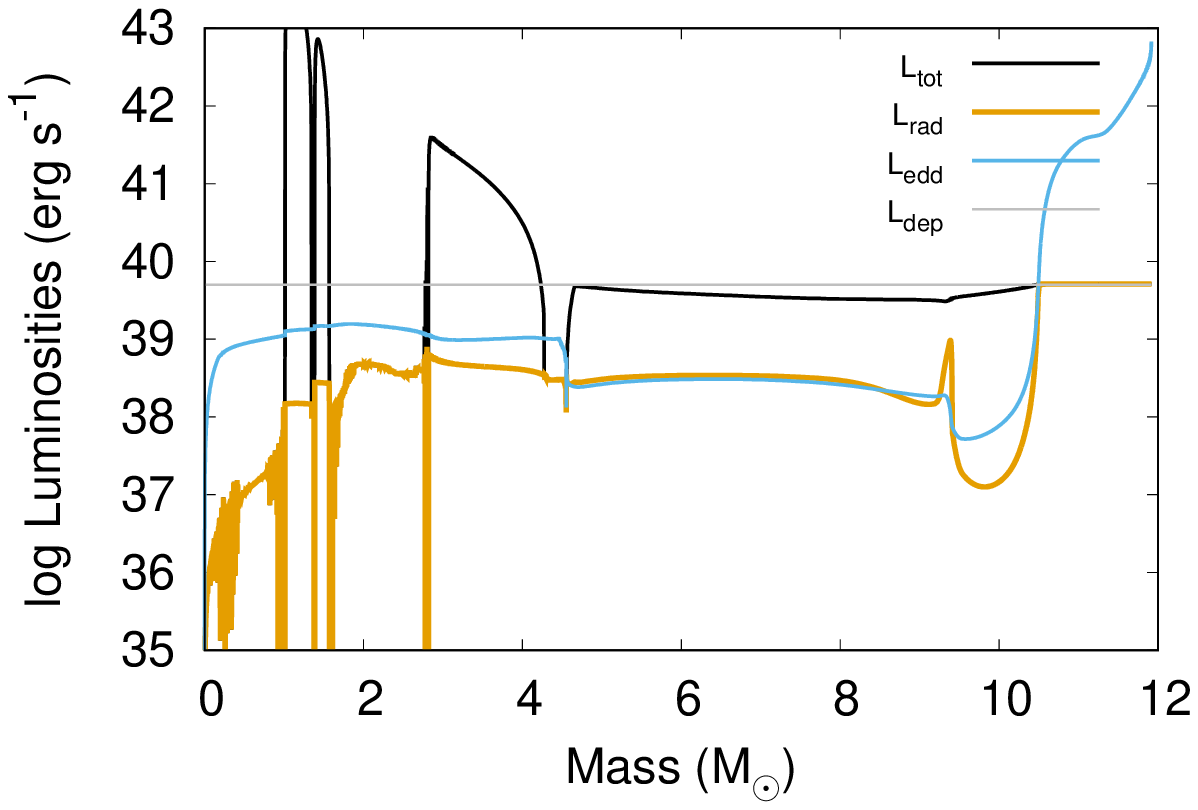}
    \end{center}
  \end{minipage}
  \end{tabular}
 \caption{Left: Total (black), radiative (orange), and Eddington (blue) luminosities as a function of mass coordinate for the progenitor at 3.0 years before the core collapse. Right: The same plot as the left panel for the model of Ldep5d39BASE at the time of oxygen exhaustion (or 34 days before the core collapse). The injection rate of $L_{\mathrm{dep}}=5.0 \times 10^{39}$ erg s$^{-1}$ is also plotted with a gray line.}
  \label{Ledd}
\end{figure*}

Fig.\ref{density_profile_different_rate} compares the density profile of the progenitors at the time of core collapse for the models with different energy injection rates. For reference, the density profile for a constant mass flux of $\dot{M} = 10^{-1} M_{\odot} \mathrm{yr}^{-1}$, assuming the constant wind velocity of $v = 10$ km s$^{-1}$, is also shown. The higher the energy injection rate is, the farther the envelope expands for a given duration (here, 3.0 years). This can be seen by inspecting the Fig. \ref{density_profile_different_rate} and outermost radius of the models ($R_{\mathrm{out}}$) shown in the Table \ref{table}. Note, however, that the photospheric radius of the model Ldep1d40uni is smaller than that of Ldep5d39uni, which seems to be opposite to the intuition (see Table \ref{table}). This happens because, especially for the former model, the majority of the hydrogen envelope is recombined, and thus the photosphere recedes in radius.

The qualitative behaviour of the different models can be understood in terms of the ratio of the injection rate to the Eddington luminosity of the progenitor. The left panel of Fig.\ref{Ledd} shows the Eddington luminosity ($L_{\mathrm{EDD}}=4 \pi G c m(r)/ \kappa (r)$) of the progenitor's envelope at 3.0 years before the core collapse, together with the total luminosity and radiative luminosity. As is clear from the figure, the Eddington luminosity in the envelope is $L_{\mathrm{EDD}} \sim 10^{38}$ erg s$^{-1}$.
When the injection rate is roughly comparable or less than the Eddington luminosity ($L_{\mathrm{dep}} \lesssim 10^{39} $ erg s$^{-1}$), the resulting radiative luminosity is below the Eddington luminosity ($10^{38}$ erg s$^{-1}$). Therefore, the envelope expands quasi-hydrostatically, and it does not expand so much in a timescale of a few years. Actually, the timescale for the readjustment of the stellar structure can be estimated as $t_{\mathrm{thermal}} \simeq \frac{G M M_{\mathrm{env}}}{2 R L} \approx 30$ yr, assuming $L=10^5 L_{\odot}$. This timescale is an order of magnitude longer than the timescale we consider in this paper.


On the contrary, when the injection rate is highly super-Eddington ($L_{\mathrm{dep}} \gg  L_{\mathrm{EDD}} \sim 10^{38}$ erg s$^{-1}$), such as the case of $L_{\mathrm{dep}} = 5.0 \times 10^{39} $ erg s$^{-1}$ and $10^{40}$ erg s$^{-1}$ , the radiative luminosity reaches the Eddington luminosity. Thus, the envelope is accelerated dynamically and it expands significantly, approaching to $\sim 10^{15}$ cm within a few years.

\subsubsection{Light curves and photospheric velocity} \label{sec_LC_vph_different_rate}

\begin{figure*}[htbp]
    \begin{tabular}{c}
    \begin{minipage}{0.5\hsize}
    \begin{center}
       \includegraphics[width=90mm, bb =50 50 410 300]{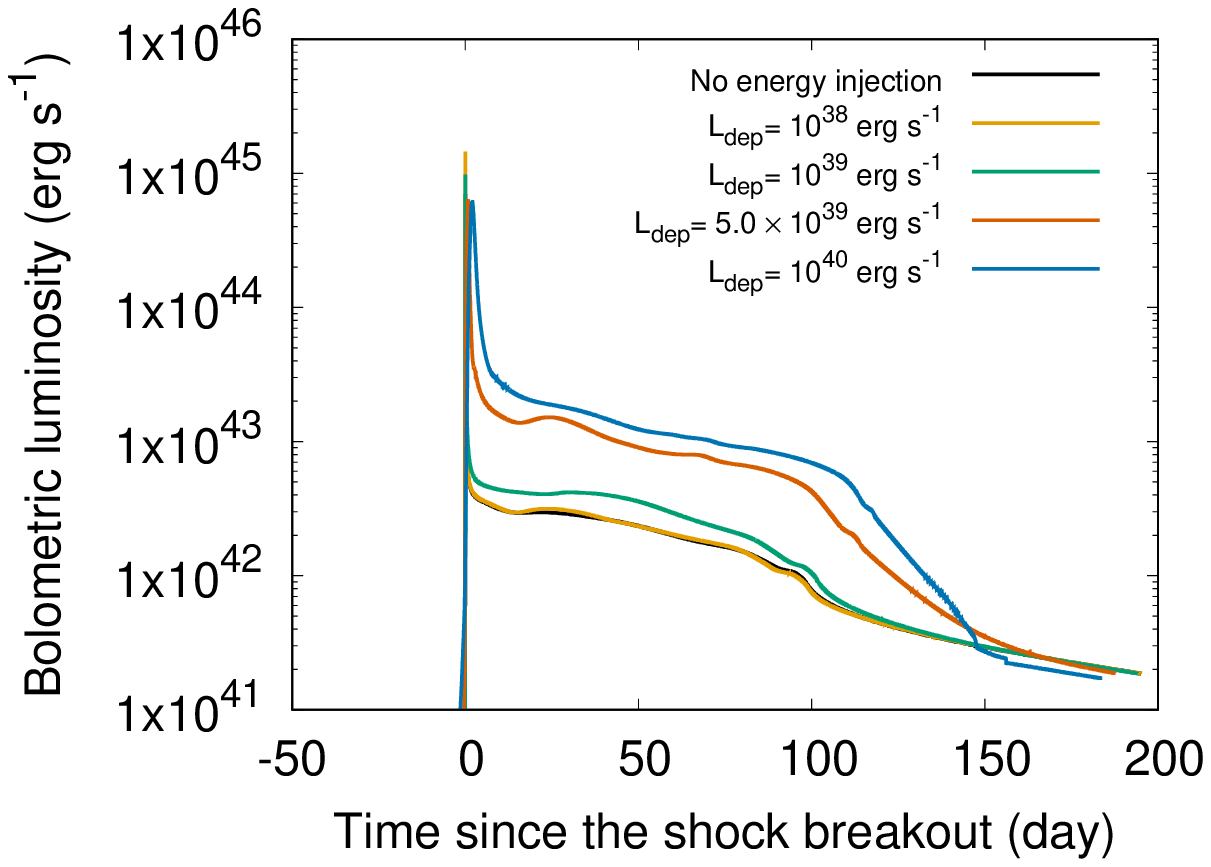}
    \end{center}
  \end{minipage}
  \begin{minipage}{0.5\hsize}
    \begin{center}
       \includegraphics[width=90mm, bb =50 50 410 300]{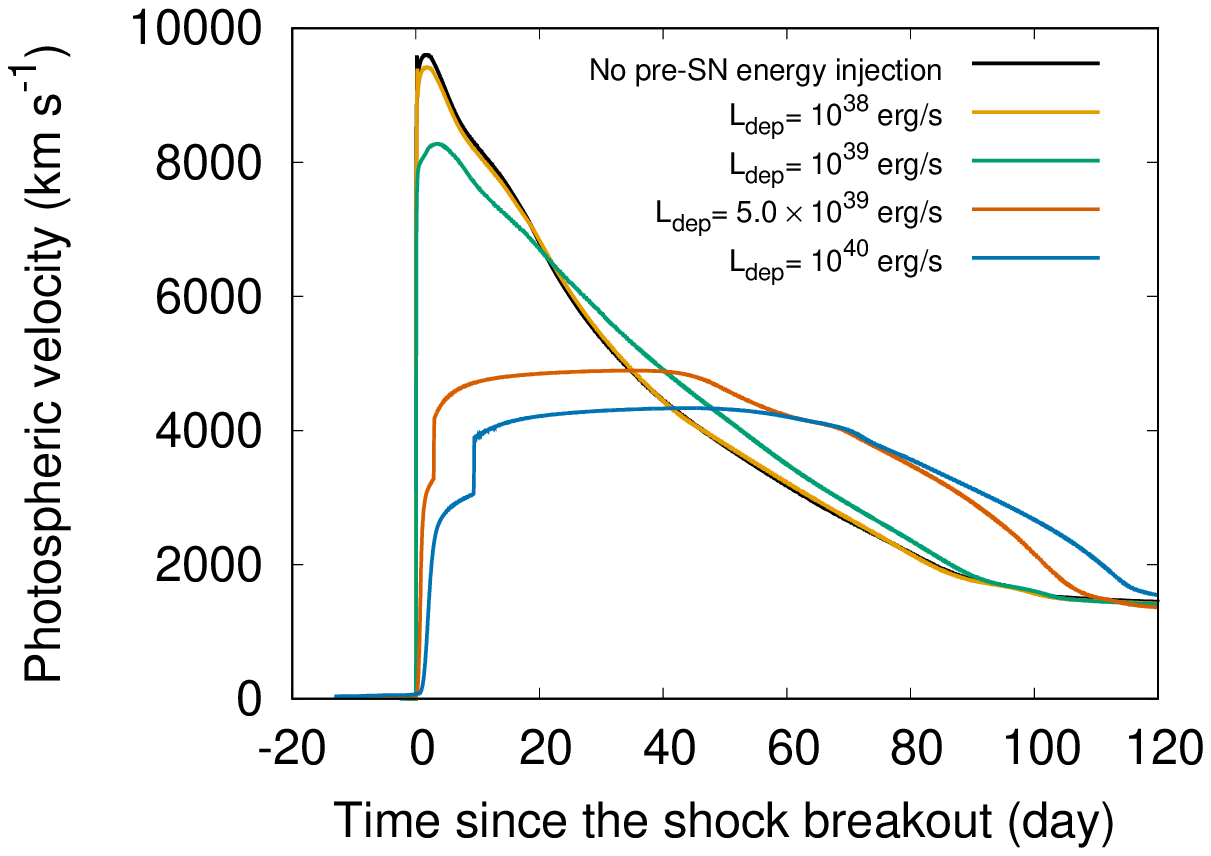}
    \end{center}
  \end{minipage}
  \end{tabular}
 \caption{Left: The bolometric light curve of the model with different rates for the energy injection. We assume UNIFORM distribution for all the models. A black line shows the model without the pre-SN energy injection. Orange, green, red and blue lines show the model with $L_{\mathrm{dep}}= 10^{38} $ erg s$^{-1}$ $, 10^{39} $ erg s$^{-1}$ $, 5.0 \times 10^{39} $ erg s$^{-1}$ $, 10^{40} $ erg s$^{-1}$, respectively. Right: The photospheric velocity of the same models as left panel, with the corresponding colors.}
  \label{LC_vph_different_rate}
\end{figure*}

The left panel of Fig.\ref{LC_vph_different_rate} compares the bolometric light curves of the models with different energy injection rates.
The higher the energy injection rate is, the later it reaches the maximum brightness since the explosion (Table\ref{table}). As explained in \S \ref{rho_different_rate}, the envelope expands farther with the higher energy injection rates, thus, the shock has to travel longer distance in order for the shock energy to diffuse out. Also, the plateau is brighter and longer for the higher energy injection rate. This can be explained by the longer expansion timescale for the models with larger initial radius (see \S \ref{sec_LC_Ldep5d39uni}). When $L_{\mathrm{dep}} \lesssim 10^{39} $ erg s$^{-1}$, the luminosity and duration of the plateau phase do not differ significantly from the model without the pre-SN energy injection. The difference is less than an order of magnitude. On the contrary, when $L_{\mathrm{dep}} \gg L_{\mathrm{EDD}} \sim 10^{38} $ erg s$^{-1}$, the effect on the light curve is significant, making the SN an order of magnitude brighter. 


The right panel of Fig.\ref{LC_vph_different_rate} compares the photospheric velocity of the models with different energy injection rates. Just like the light curves, when $L_{\mathrm{dep}} \lesssim 10^{38}$ erg s$^{-1}$, the velocity evolution does not deviate from the model without the pre-SN energy injection so much. When $L_{\mathrm{dep}} \gg L_{\mathrm{EDD}}$, on the other hand, the velocity evolution is completely different, being relatively constant until $t \sim 50$ days. 

To summarize, when $L_{\mathrm{dep}} \gg L_{\mathrm{EDD}} \sim 10^{38}$ erg s$^{-1}$, such as the case of $L_{\mathrm{dep}} = 5.0 \times 10^{39} $ erg s$^{-1}$ and $10^{40}$ erg s$^{-1}$, both the light curve and the photospheric velocity of the SN become inconsistent with the observational data for SNe II (see, Fig.\ref{LC_Ldep5d39uni}, Fig.\ref{vph_and_Rph_Ldep5d39}). Thus, in reality, such a highly super-Eddington energy injection, continuing for the last few years in the massive star evolution should not be realized in most of the SN II progenitors. On the contrary, if the injection rate is roughly comparable or less than the Eddington luminosity ($L_{\mathrm{dep}} \lesssim 10^{39} $ erg s$^{-1}$), such as the case of $L_{\mathrm{dep}} = 10^{38} $ erg s$^{-1}$ and $10^{39}$ erg s$^{-1}$, the effect on the light curves and photospheric velocity is not significant. Therefore, such an energy injection is allowed. The two distinct behaviour of our models depending on the ratio of $L_{\mathrm{dep}}$ to $L_{\mathrm{EDD}}$ is also seen in the unbound mass of the progenitor for each model (see Table\ref{table}). For the models with $L_{\mathrm{dep}} \lesssim 10^{39}$ erg s$^{-1}$, the progenitor is completely bound, while for the models with $L_{\mathrm{dep}} \gg L_{\mathrm{EDD}} \sim 10^{38} $ erg s$^{-1}$, almost entire envelope becomes unbound. Here, the unbound mass for the model Ldep1d40uni is a little smaller than the one for the model Ldep5d39uni. Especially, for the former model, the radiative luminosity becomes so large due to the recombination of the hydrogen in the the envelope during the energy injection, so that a part of the envelope loses the energy to the radiation.

Here, we have interpreted the behaviours seen in the different models by comparing the injection rate to the Eddington luminosity of the progenitor. The luminosity of the progenitor at 3.0 years before the core collapse is $L = 3 \times 10^{38}$ erg s$^{-1}$, which is indeed comparable to the Eddington luminosity (see Table\ref{table} and Fig.\ref{Ledd}). Thus, the ratio of the injection rate to the progenitor's luminosity can be practically used as a measure of the strength of the effect. However, we conclude that adopting the ratio of the injection rate to the Eddington luminosity of the progenitor is a straightforward criterion which physically divides the behaviors seen in the different models. The right panel of Fig.\ref{Ledd} shows that $L_{\mathrm{rad}} \gtrsim L_{\mathrm{EDD}}$ in the major part of the envelope for the model of Ldep5d39uni, indicating that radiation pressure is accelerating the envelope overcoming its self-gravity. The injection rate of $L_{\mathrm{dep}} = 5.0 \times 10^{39}$ erg s$^{-1}$ is much higher than the Eddington luminosity $L_{\mathrm{EDD}} \sim 10^{38}$ erg s$^{-1}$, thus the radiative luminosity reaches the Eddington luminosity.
Note also that, in the region where $L_{\mathrm{rad}} \gtrsim L_{\mathrm{EDD}}$, the ratio of $L_{\mathrm{rad}}$ to $L_{\mathrm{EDD}}$ is close to 1. This behaviour is the natural consequence for a star which has the near-Eddington luminosity \citep{2006A&A...450..219P, 2012A&A...538A..40G}.


\subsection{Evolution of the progenitor on the HR diagram} \label{sec_HR}

\begin{figure*}[htbp]
    \begin{tabular}{c}
    \begin{minipage}{0.5\hsize}
    \begin{center}
       \includegraphics[width=90mm, bb =50 50 410 300]{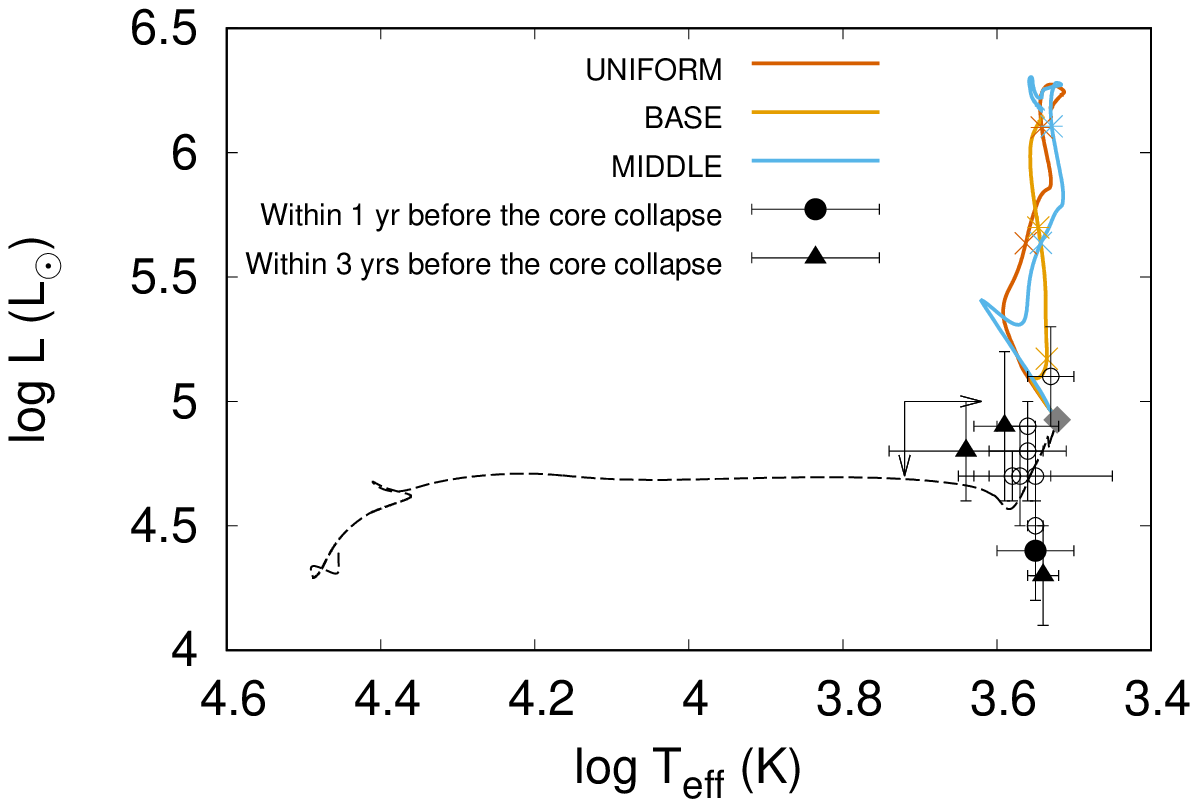}
    \end{center}
  \end{minipage}
  \begin{minipage}{0.5\hsize}
    \begin{center}
      \includegraphics[width=90mm, bb =50 50 410 300]{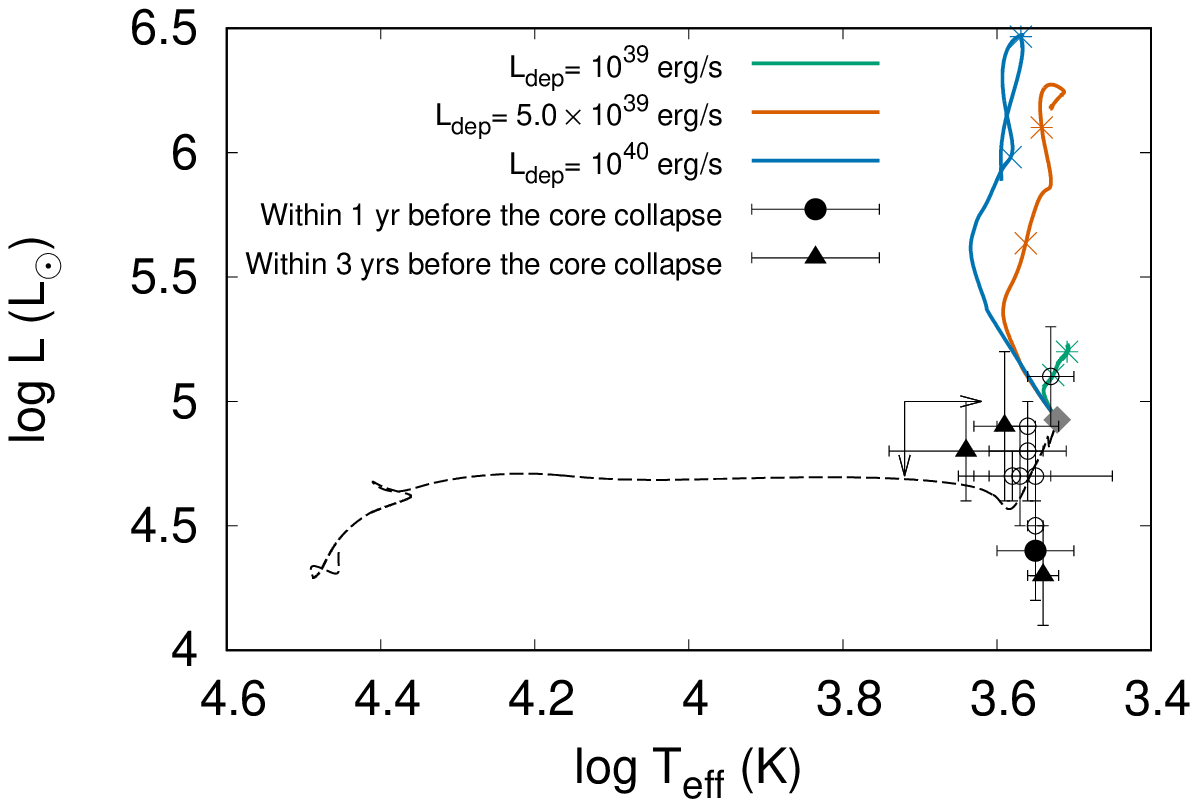}
    \end{center}
  \end{minipage}
  \end{tabular}
  \caption{Left: The evolution of the progenitor on the HR-diagram for the models with different energy injection locations. Here, the luminosity and stellar radius are taken at the photosphere, and not at the outermost cell. A black dotted line shows the model without the pre-SN energy injection from just before the main sequence until the core collapse. Red, orange, and blue lines show the models with UNIFORM, BASE, and MIDDLE distribution with $L_{\mathrm{dep}}=  5.0 \times 10^{39} $ erg s$^{-1}$ fixed, respectively. For each model, the locations at 1yr, 2yr before the core collapse are denoted by asterisks and crosses, respectively. The observational data for the detected progenitors of SNe IIP/IIL are also plotted. 
The data shown with filled circles and filled triangles are the data which were taken within 1 yr and 3 yr before the estimated explosion date of SN, respectively. Other data are shown with open circles. The data are taken from \citet{2015PASA...32...16S} and the references therein. Right: The evolution of the progenitor on the HR-diagram for the models with different energy injection rates. We assume UNIFORM distribution. A black dotted line shows the model without the pre-SN energy injection. Green, red, and blue lines show the models with $L_{\mathrm{dep}}= 10^{39} $ erg s$^{-1}$, $5.0 \times 10^{39}$ erg s$^{-1}$, $10^{40}$ erg s$^{-1}$, respectively. Here, the data with $L_{\mathrm{dep}}= 10^{38} $ erg s$^{-1}$ shows no noticeable movement on the plot, so we simply did not show it. The observational data are the same as the left panel. For both panels, the location of the progenitor at the time of core collapse when there is no pre-SN energy injection is denoted by a filled gray diamond.}
  \label{HR}
\end{figure*}

The left panel of Fig.\ref{HR} shows the evolution of the progenitors on the HR diagram for the models with different energy injection locations, with $L_{\mathrm{dep}}=  5.0 \times 10^{39} $ erg s$^{-1}$ fixed. The observational data for the detected progenitors of SNe IIP/IIL \citep{2015PASA...32...16S} are also plotted. For all the models, the luminosity increases rapidly within a timescale of a few years, finally reaching $L \gtrsim 10^6 L_{\odot}$. The time evolution of luminosity is different for the three models, in a sense that the BASE model becomes bright later than the other two. 



The right panel of Fig.\ref{HR} shows the evolution of the progenitors on the HR diagram for the models with different energy injection rates, with the deposition location fixed as UNIFORM. For the model with $L_{\mathrm{dep}} = 10^{38}$ erg s$^{-1}$ or $10^{39}$ erg s$^{-1}$, the location on the HR diagram changes little for 3 years. Thus, these models are consistent with the properties of the detected progenitors of SNe II.

On the contrary, for the other two models with highly super-Eddington luminosity ($L_{\mathrm{dep}} \gtrsim 5 \times 10^{39}$ erg s$^{-1}$), the progenitor luminosity reaches $\gtrsim 4 \times 10^5 L_{\odot}$ within $\sim 1$ yr since the energy injection starts. Such a high luminosity is inconsistent with the properties of the detected progenitors of SNe II. Furthermore, for some SNe IIP, the existence of significant variability has been ruled out during the last yr -- decades of their lives \citep{2017MNRAS.467.3347K, 2018MNRAS.480.1696J}. Thus, such a high energy injection rate ($L_{\mathrm{dep}} \gtrsim 5 \times 10^{39}$ erg s$^{-1}$) also contradicts with these observations. These findings support our conclusion derived in \S \ref{sec_different_rate} that highly super-Eddington energy injection ($L_{\mathrm{dep}} \gg L_{\mathrm{EDD}} \sim 10^{38} $ erg s$^{-1}$) continuing for a few years before the core collapse should not be realized for most of the SNe II progenitors.



\section{Discussion} \label{discussion}

\subsection{Comparison with analytic scaling relations} 
 
\begin{figure*}[t]
    \begin{tabular}{c}
    \begin{minipage}{0.5\hsize}
    \begin{center}
       \includegraphics[width=90mm, bb =50 50 410 300]{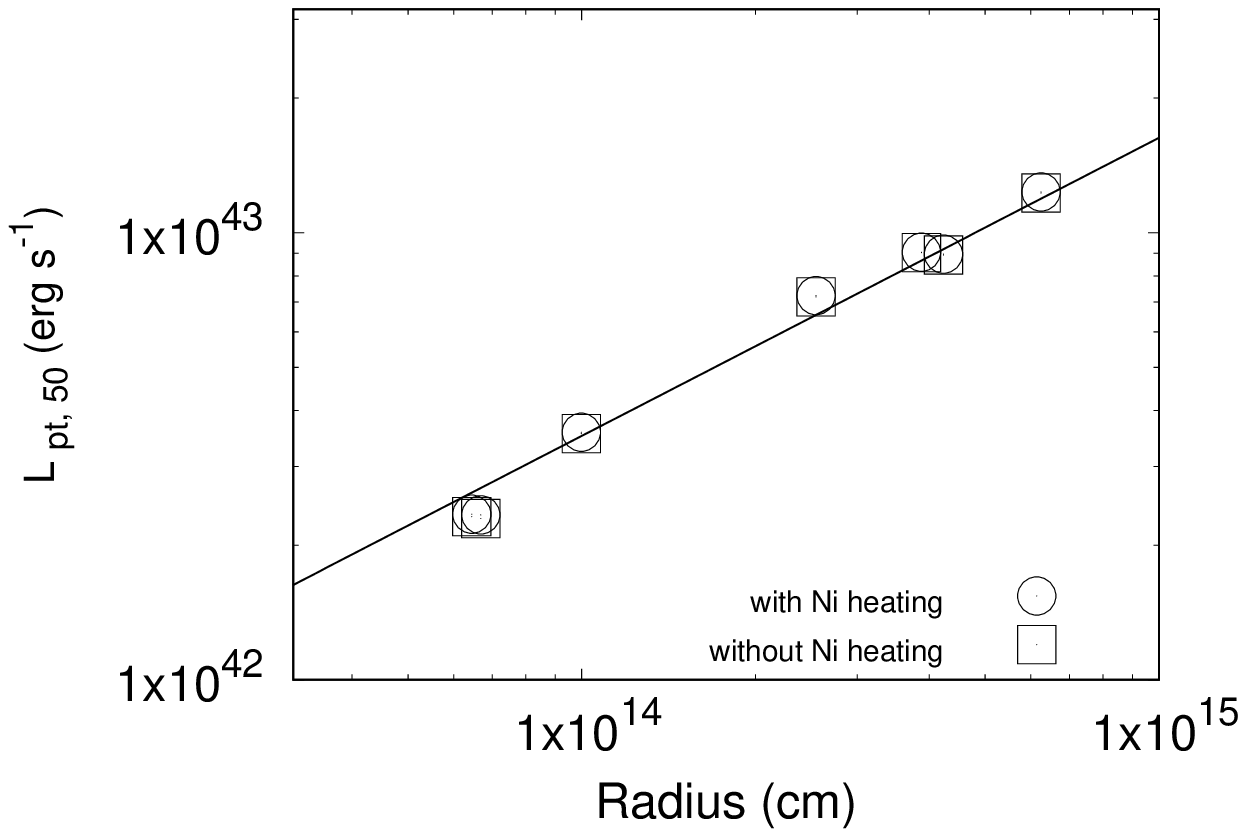}
    \end{center}
  \end{minipage}
  \begin{minipage}{0.5\hsize}
    \begin{center}
       \includegraphics[width=90mm, bb =50 50 410 300]{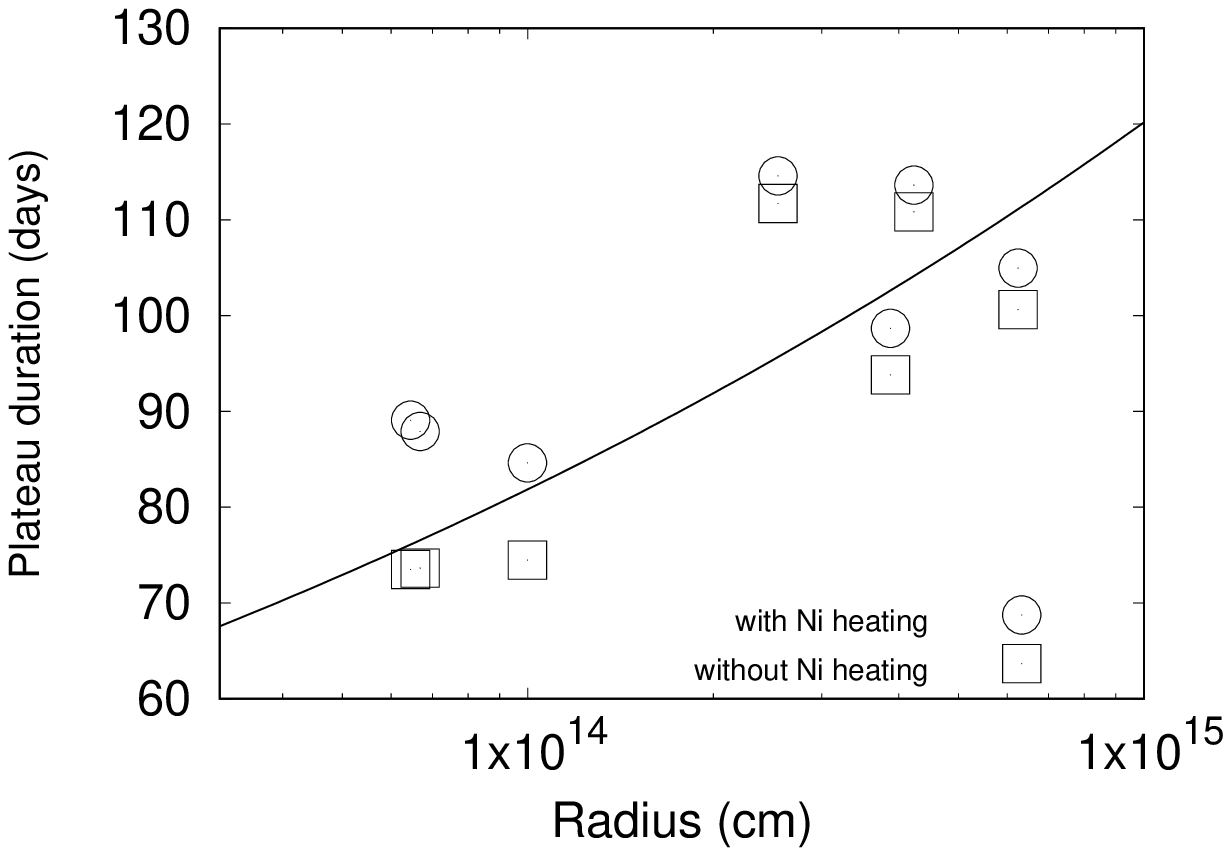}
    \end{center}
  \end{minipage}
  \end{tabular}
 \caption{Left: The plateau luminosity ($L_{\mathrm{pt}, 50}$) of our model sequence as compared with the analytic scaling relations \citep{1993ApJ...414..712P, 2009ApJ...703.2205K}. Here, the plateau luminosity is represented by the value at 50 days since the shock breakout. A line in the plot denotes $L_{\mathrm{pt}, 50} \propto R^{2/3}$, with an arbitrary normalization. The open circles denote all the models shown in Table\ref{table}. The open squares denote the same models, but switching off the $^{56}$Ni heating. Right:  The plateau duration of our model sequence as compared with the analytic scaling relation \citep{1993ApJ...414..712P, 2009ApJ...703.2205K}. The plateau duration of the models is calculated as the time interval between the shock breakout and the time when the luminosity goes down to the half of $L_{\mathrm{pt}, 50}$. A line in the plot denotes $t_{\mathrm{SN}} \propto R^{2/3}$, with an arbitrary normalization. The plotted models are the same as the left panel. Note that for both panels, the radius in the $x$-axis is the value at the outer most cell of the model, and not at the photosphere.}
  \label{compare_with_analytics}
\end{figure*}

In this section, we compare our results with the analytic scaling relations for SNe IIP derived by \citet{1993ApJ...414..712P} and \citet{2009ApJ...703.2205K}:
  \begin{eqnarray}
    L_{\mathrm{SN}} \propto E^{5/6} M_{\mathrm{ej}}^{-1/2} R_0^{2/3}\kappa^{-1/3}T_{\mathrm{I}}^{4/3}, \\
    t_{\mathrm{SN}} \propto E^{-1/6} M_{\mathrm{ej}}^{1/2} R_0^{1/6} \kappa^{1/6} T_{I}^{-2/3},
    \end{eqnarray}
 where $L_{\mathrm{SN}}$ and $t_{\mathrm{SN}}$ denote the luminosity and duration of the plateau phase. Here, $E$ is the explosion energy, $M_{\mathrm{ej}}$ is the ejecta mass, $R_0$ is the pre-supernova progenitor radius and $\kappa$, $T_{\mathrm{I}}$ are opacity and ionization temperature, respectively. 
  Note that these relations are derived under the assumption that the energy source is solely comes from the explosion energy. Thus, the $^{56}$Ni heating is not taken into account in these relations.

The left panel of Fig.\ref{compare_with_analytics} compares the plateau luminosity of our model sequence shown in Table\ref{table} to the analytic scaling relation. Here, we adopt the analytic scaling relation as $L_{\mathrm{SN}} \propto R_0^{2/3}$, since other parameters such as $E$, $M_{\mathrm{ej}}$, $\kappa$, and $T_{\mathrm{I}}$ are considered the same for the different models. 
  For the comparison purpose, we have additionally run the models including no $^{56}$Ni heating, which are also shown in Fig.\ref{compare_with_analytics}. The existence of the $^{56}$Ni heating does not cause the noticeable difference in the plateau luminosity. This is because at 50 days since the shock breakout, the photosphere has not yet been receded to the inner region which has been heated by $^{56}$Ni. The plateau luminosity of our model sequence is well fit by the analytic scaling relation. As has been explained in the \S \ref{sec_LC_vph_different_rate}, the plateau luminosity is higher for the larger progenitors, because of the longer expansion timescale.

The right panel of Fig.\ref{compare_with_analytics} compares the plateau duration of our model sequence shown in Table 1 to the analytic scaling relation. Here, we adopt the analytic scaling relation as  $t_{\mathrm{SN}} \propto R_0^{1/6}$, assuming the other parameters are the same for the different models. In contrast to the plateau luminosity, the existence of $^{56}$Ni heating makes the plateau duration longer, especially for the models with smaller progenitor radius. This is because the heating by $^{56}$Ni slows down the cooling of the envelope, thus delaying the ionization front to reach the core. This effect is less significant for the initially larger progenitors, because they retain larger internal energy and the relative contribution of the $^{56}$Ni heating is minor. The analytic scaling relation matches our model sequence reasonably well, if the effect of the $^{56}$Ni heating is omitted. As has been explained in the \S \ref{sec_LC_vph_different_rate}, the larger progenitors have longer plateau because they have longer expansion timescale.


Thus, both the luminosity and duration of the plateau phase for our model sequence are well fit with the analytic relations derived for SNe IIP. This indicates that the interpretation of our models as an ``expanded envelope'' explains not only the slow evolution of temperature (Fig.\ref{Tph_Ldep5d39} and Fig.\ref{T_evolution}), and longer time from the explosion to the shock breakout (Table \ref{table}), but also the different plateau luminosity and duration.

\subsection{Mechanisms of the pre-SN mass loss}

We have shown that the highly super-Eddington energy injection ($L_{\mathrm{dep}} \gg L_{\mathrm{EDD}} \sim 10^{38}$ erg s$^{-1}$), continuing for a few year before the core collapse, should not be realized in most of the SNe II. On the contrary, if the energy injection rate is roughly comparable or less than the Eddington luminosity ($L_{\mathrm{dep}} \lesssim 10^{39} $ erg s$^{-1}$), it is difficult to produce the CSM within a few years with $\dot{M} \approx 10^{-3} M_{\odot} \mathrm{yr}^{-1}$, extending to $\sim 10^{14} - 10^{15}$ cm, which is inferred for SNe II \citep{2017NatPh..13..510Y, 2018NatAs.tmp..122F}. The quasi-hydrostatic expansion is slow with the timescale of $\gtrsim 10$ years, and the envelope does not expand thus far, within a few years. Furthermore, even if the sub-Eddington energy injection continues much longer than a few years and expands to $\gtrsim 10^{14}$ cm, the envelope is expected to be too dense so that it will behave like an extremely extended red supergiant, which contradicts with the observations. From these arguments, it is unlikely that energy injection directly triggers the mass loss which is responsible for the confined CSM. 

Rather, we propose a scenario that the modest envelope inflation, which is caused by the sub-Eddington energy injection, triggers the secondary effects. One possibility is stellar pulsation. Actually, \citet{2010ApJ...717L..62Y} have shown that the pulsation growth rate increases as the stellar radius increases.
Thus, it is possible that the envelope inflation, caused by the moderate energy injection amplifies the stellar pulsation and thus, induces the enhanced mass loss. In order for this to work, the energy injection should start from $\gtrsim 10 $ years before the core collapse, because the envelope expands with the thermal timescale of the envelope. Under the hypothesis of the energy injection by gravity waves, this may require that energy injection should be significant already from the carbon shell burning stage \citep{2017MNRAS.467.3347K}.


Another possibility of the secondary effect can be a binary interaction. When the progenitor is close to fill the Roche-lobe, the envelope inflation caused by the sub-Eddington energy injection triggers the Roche-lobe overflow \citep{2014MNRAS.445.2492M}. In this case, a fraction of the transferred mass might be driven out of the system, constructing a dense CSM \citep{2005A&A...435.1013P, 2017ApJ...840...90O}. In an extreme case, it may lead to the common envelope interaction, which ejects the envelope by using the orbital energy \citep{1976IAUS...73...75P, 2013A&ARv..21...59I}. We will examine these different scenarios further in a forthcoming paper (Ouchi et al., in prep.).

The arguments above are based on the results of our simulations. We note that there are a few caveats. Firstly, our model sequence does not cover the situation where the energy injection takes place near the stellar surface. However, we note that the energy injection likely takes place deep in the envelope, if this is related to an additional energy generation in the core; for example, the estimations by \citet{2012MNRAS.423L..92Q} and \citet{2017MNRAS.470.1642F} show that the dissipation of waves via radiative diffusion is most likely to take place at the relatively inner part of the envelope ($r \sim 30$ -- $100 R_{\odot}$), or near the base of hydrogen envelope, respectively. Furthermore, there are several works that have investigated the response of the envelope to the near-surface energy deposition, using similar approach to ours \citep{2014MNRAS.445.2492M, 2016MNRAS.458.1214Q}, and their results are qualitatively similar to ours.

Secondly, there is an uncertainty in the progenitor's exact density structure in the optically thin region above the photosphere, due to the simple atmosphere model we have used (see Appendix \ref{appendix_A}). However, we conclude that it would not affect our result significantly for the following reason. Fig.\ref{density_profile_different_locations} and Fig.\ref{LC_vph_different_location} show that the light curve and photospheric velocity behave quite similarly for the three models with different locations for the energy injection, although the density structures above the photosphere are quite different. The difference in the structure is large, and it may well cover the difference introduced by uncertainties introduced by the atmospheric model.

\subsection{Possible applications to peculiar SNe}

In this section, we summarize the observational characteristics, expected for the SNe from the progenitors with pre-SN energy injection, and discuss its properties in relation to some observed objects. In the case when the injection rate is highly super-Eddington ($L_{\mathrm{dep}} \gg L_{\mathrm{EDD}} \sim 10^{38}$ erg s$^{-1}$), the envelope expands to $\gtrsim 10^{14}$ cm within a few years, and both light curve and photospheric velocity show peculiar evolution. The resulting SNe will look like SNe IIP with a plateau. However, it takes $\gtrsim 10$ days from the time of explosion to reach the maximum brightness, which is much longer than the model without the pre-SN energy injection. Moreover, the plateau luminosity is an order of magnitude brighter than the model without the pre-SN energy injection. The plateau is long and continues until $\gtrsim 100$ days after explosion, which is unusually longer than the observed SNe II. Photospheric velocity is relatively constant for more than $\sim$ 50 days, and after that, it begins to decline.


The plateau-type SN2009kf has unusually bright bolometric and near-ultraviolet luminosity and high velocity at late times \citep{2010ApJ...717L..52B}. Therefore, the extremely high explosion energy of $E_{\mathrm{exp}} =2.2 \times 10^{52}$ erg, together with the large ejecta mass of $M_{\mathrm{ej}} = 28.1 M_{\odot}$ have been suspected in order to fit both the light curves and the expansion velocity \citep{2010ApJ...723L..89U}. Our models with super-Eddington energy injection rate can explain these observational data without assuming extremely high explosion energy (Ouchi et al., in prep).  Therefore, such a super-Eddington energy injection might be realized in some stars. However, it should be rare, considering the fraction of such peculiar objects in the observed samples. Note also that \citet{2011MNRAS.415..199M} have proposed that the observational data of SN2009kf can be well fit by attaching the dense CSM ($\dot{M} = 10^{-2} M_{\odot}$ yr$^{-1}$) above the stellar surface.

Several SNe IIP, such as SN 2009ib, and 2015ba, are known to have unusually long plateau, lasting for $\gtrsim 120$ days since the explosion. This feature is also consistent with our models with super-Eddington energy injection. However, the velocity evolution of these SNe are quite typical for SNe II, declining rapidly soon after the explosion \citep{2015MNRAS.450.3137T, 2018MNRAS.479.2421D}. Thus, our models seem not to be applicable to those SNe.


If the energy injection rate is roughly comparable or less than the Eddington luminosity, $L_{\mathrm{dep}} \lesssim 10^{39} $ erg s$^{-1}$, then the envelope expands only to $\lesssim 10^{14}$ cm, and the plateau luminosity is higher than the typical SNe II only by a factor of a few. The velocity decline rate is slower than the model without the pre-SN energy injection. The effect on the light curves and velocity evolution of SNe, however, are not significant, and it might be difficult to distinguish the model with sub-Eddington energy injection from the model without it. However, it may be possible to find its effect on the time variations of the luminosity of the progenitors, once a large sample of SN II progenitors are analyzed in a statistical way (Fig.\ref{HR}).


\subsection{Application to the theory of gravity waves} 

Here, we consider the application of our results to the hypothesis of gravity waves. Several works have been done to estimate the wave energy deposition rate into the envelope, and investigate its effect on the structure of the envelope \citep{2012MNRAS.423L..92Q, 2017MNRAS.470.1642F, 2018MNRAS.476.1853F}. Note, however, that these models suffer various uncertainties which are difficult to evaluate from the first principal. It has been estimated that the wave energy deposition rate during the core neon burning and core oxygen burning stages can be highly super-Eddington, possibly reaching $L_{\mathrm{dep}} = 10^{41}$ -- $10^{42}$ erg s$^{-1}$ \citep{2012MNRAS.423L..92Q, 2014ApJ...780...96S}. The calculation by \citet{2017MNRAS.470.1642F} for the model of $M_{\mathrm{ZAMS}} = 15 M_{\odot}$ shows that the energy injection rate due to the gravity waves becomes highly super-Eddington ($\gtrsim 10^{40}$ erg s$^{-1}$) during the core oxygen burning, which takes place $\sim 0.5$ yr before the core collapse for their model. Because this duration is shorter than what we considered in this paper, the envelope expands only to $\sim 1750 R_{\odot}$ in their model, which is shorter than our highly super-Eddington models. Rather, this result is similar to our model with the moderate energy deposition of $L_{\mathrm{dep}}= 10^{39}$ erg s$^{-1}$, which continues for 3 years. Therefore, it should behave like 
a relatively expanded envelope, and we expect that the effect on the light curves and photospheric velocity of SNe are not significant.

However, the timescales of each burning stage are different between the stars, depending primarily on the star's mass and metalicity. Actually, the timescale of both the convective neon and oxygen burning phase, during which the wave energy deposition rate is expected to be highly super-Eddington, can be as long as $\gtrsim 5$ yr, depending on the models \citep{2002RvMP...74.1015W, 2014ApJ...780...96S}. We propose that such a super-Eddington energy injection should not be realized in most of the SNe II progenitors. This might indicates limitations in some treatments in the gravity wave models.
The fraction of wave energy that penetrate into the envelope may have simply been overestimated.
Also, the change of the density structure by the energy deposition might suppress the fraction of the wave power that can tunnel to the envelope \Citep{2012MNRAS.423L..92Q}.
These arguments highlight the importance of our approach, calibrating the models with the observational data, in constraining the uncertain physical processes involved in the progenitor models.

\section{Conclusions} \label{conclusion} 
Evidence has been accumulating that some massive stars experience the enhanced mass loss just prior (years - decades) to the SNe explosion. The physical mechanism for this phenomenon has not been clarified. Energy deposition into the envelope by certain mechanisms like gravity waves or binary interaction has been proposed as a possible cause of mass loss.

In this work, we have investigated the response of the envelope to various kinds of sustained energy deposition, which starts from a few years before the core collapse. We also calculated the effect of them on the light curves of SNe, self-consistently. We have found that the envelope expansion is triggered by the energy injection, so that the SNe experience the shock breakout in an expanded envelope.

We have also found that if highly super-Eddington energy deposition takes place, which exceeds the Eddington luminosity by more than an order of magnitude, the location of the progenitor on the HR diagram, the light curves and the evolution of photospheric velocity are all inconsistent with the observations of SNe II. Thus, we conclude that such a highly super-Eddington energy injection continuing the last few years should not be realized in most of the SNe II progenitors.

On the contrary, if the energy injection rate is moderate and does not exceed the Eddington luminosity by more than one order of magnitude, the envelope expands quasi-hydrostatically with the thermal timescale of $\gtrsim 10$ years. Therefore, the effect of the energy injection on the progenitor and SN is not significant. However, with such energy injection, it is difficult to produce the extended CSM ($\gtrsim 10^{14}$ cm) within a few years, which is inferred for SNe II. Furthermore, even if such moderate energy injection continues much longer than a few years and expands to $\gtrsim 10^{14}$ cm, the envelope is likely to be too dense so that it will behave like an extremely extended red supergiant, which contradicts with the observations. From these, it is unlikely that energy injection directly triggers the mass loss to create the observationally inferred dense and confined CSM.

As an alternative scenario, we propose a hypothesis that a secondary effect triggered by the moderate envelope inflation, which is caused by
the sub-Eddington energy injection, likely induces the mass loss. Candidates include stellar pulsation and binary interaction, which we will investigate in the future (Ouchi et al., in prep).

\acknowledgments We thank the anonymous referee for the constructive and useful comments that helped improve this manuscript.
We also thank Francisco F{\"o}rster, Takashi Moriya, Nozomu Tominaga, Masaomi Tanaka, Akihiro Suzuki, and Takashi Nagao for useful discussion. This research was supported by a grant from the Hayakawa Satio Fund 
awarded by the Astronomical Society of Japan. K.M. acknowledges support from JSPS Kakenhi grants (18H05223, 18H04585 and 17H02864).

\software{MESA \citep{2011ApJS..192....3P, 2013ApJS..208....4P, 2015ApJS..220...15P, 2018ApJS..234...34P}, SNEC \citep{2015ApJ...814...63M}
          }

\appendix

\section{STELLAR EVOLUTION USING MESA} 

\subsection{Evolving to He exhaustion}  \label{appendix_A}

For the calculation of stellar evolution, we use the one-dimensional stellar evolution code MESA of version 10398 \citep{2011ApJS..192....3P, 2013ApJS..208....4P, 2015ApJS..220...15P, 2018ApJS..234...34P}. The procedure of the calculation closely follows \citet{2017MNRAS.470.1642F}, although there are several differences, in addition to the version of MESA which is used.

First, we evolve a $15M_{\odot}$ non-rotating star from pre-main sequence to the exhaustion of He, assuming hydrostatic:

   \texttt{change\_initial\_v\_flag = .true.}

   \texttt{new\_v\_flag = .false.} \\
   We use the default parameter settings of massive stars for most of the parameters, with several modifications. The initial metalicity is assumed to be Z=0.02. For the atmosphere model, we adopt
   
  \texttt{which\_atm\_option = 'simple\_photosphere'} \\
We set the mixing length parameter to be $\alpha = 1.9$, and do not use MLT++ scheme, this time. We set the parameters for overshooting as,

\texttt{overshoot\_f0\_above\_nonburn\_core = 0.004}

\texttt{overshoot\_f0\_above\_nonburn\_shell = 0.004}

\texttt{overshoot\_f0\_below\_nonburn\_shell = 0.004}
\vspace{1ex}

\texttt{overshoot\_f\_above\_nonburn\_core = 0.010}       
 
\texttt{overshoot\_f\_above\_nonburn\_shell = 0.010}         

\texttt{overshoot\_f\_below\_nonburn\_shell = 0.010}, \\
and use the same setting for H, He and Z core/shell burning. Additionally, we set,

\texttt{overshoot\_D2\_below\_burn\_z = 1d10}

\texttt{overshoot\_f2\_below\_burn\_z = 0.10}. \\
For the stellar wind, we use the Dutch scheme as follws:

\texttt{hot\_wind\_full\_on\_T = 1d0}

\texttt{cool\_wind\_full\_on\_T = 0d0}

\texttt{hot\_wind\_scheme = 'Dutch'}

\texttt{Dutch\_wind\_lowT\_scheme = 'de Jager'}

\texttt{Dutch\_scaling\_factor = 1.0}. \\

After we evolve the model to the exhaustion of He, we stop the calculation and restart with the following commands:

\texttt{relax\_initial\_tau\_factor = .true.}

\texttt{relax\_to\_this\_tau\_factor = 1d-4}

\texttt{dlogtau\_factor = 1d-1}.\\
This allows us to evolve the star above the photosphere up to the region of $\tau = 10^{-4}$. After setting this command, we evolve the star for 100 years, with a maximum timestep of 1.0 yr so that the star is close to a hydrostatic equilibrium.

\subsection{Hydrodynamic simulation until core collapse} \label{appendix_A2}

Then, we turn on the hydrodynamic mode of MESA as follows:

\texttt{change\_initial\_v\_flag = .true.}

\texttt{new\_v\_flag = .true.} \\
The hydro equations and the boundary conditions are specified with the following commands:

 \texttt{use\_ODE\_var\_eqn\_pairing = .true.}      

 \texttt{use\_dvdt\_form\_of\_momentum\_eqn = .true.}

 \texttt{use\_compression\_outer\_BC = .true.}
 
 \texttt{use\_T\_Paczynski\_outer\_BC = .true.}. \\
 We apply artificial viscosity in order to allow MESA to resolve hydrodynamic shocks with the following commands:

 \texttt{use\_artificial\_viscosity = .true.}
 
  \texttt{shock\_spread\_linear = 0}
 
  \texttt{shock\_spread\_quadratic = 1d-2.} \\
For the convection, we adopt the following limiting scheme:

 \texttt{min\_T\_for\_acceleration\_limited\_conv\_velocity = 0}

 \texttt{max\_T\_for\_acceleration\_limited\_conv\_velocity=1d11}

 \texttt{mlt\_accel\_g\_theta = 1d0.} \\
For the gridding and the error tolerances, we adopt the following values:

       \texttt{mesh\_delta\_coeff = 0.8d0}
      
      \texttt{log\_tau\_function\_weight =  100}  

      \texttt{log\_kap\_function\_weight =  100}  

      \texttt{max\_surface\_cell\_dq = 1d-10}
      \vspace{1ex}
      
   \texttt{newton\_iterations\_limit = 9} 

   \texttt{iter\_for\_resid\_tol2 = 6}

   \texttt{tol\_residual\_norm1 = 1d-8}

   \texttt{tol\_max\_residual1 = 1d-7}
 
       \texttt{tiny\_corr\_coeff\_limit = 999999}

       \texttt{newton\_itermin\_until\_reduce\_min\_corr\_coeff = 999999}.
       
       With these settings, we evolve the model until 3.0 years before the core collapse. Then, we start injecting energy with the following limitation of the time step, and evolve until the infall of iron core.

 \texttt{max\_years\_for\_timestep  = 1d-3}.

Finally, we save the final model at the time of core collapse, and use it as an input model for the radiation hydrodynamic simulation, using SNEC.

\end{document}
